\documentclass[fleqn,usenatbib]{mnras}

\usepackage{newtxtext,newtxmath}

\usepackage[T1]{fontenc}
\usepackage{ae,aecompl}

\usepackage{graphicx}	
\usepackage{amsmath}	
\usepackage{amssymb}	

\def\gs{\mathrel{\raise0.35ex\hbox{$\scriptstyle >$}\kern-0.6em
\lower0.40ex\hbox{{$\scriptstyle \sim$}}}}
\def\ls{\mathrel{\raise0.35ex\hbox{$\scriptstyle <$}\kern-0.6em
\lower0.40ex\hbox{{$\scriptstyle \sim$}}}}

\title[AS2UDS: Source catalogue and properties]{An ALMA survey of the SCUBA-2 Cosmology Legacy Survey UKIDSS/UDS field: Source catalogue and properties}

\author[Stach et al.]{
Stuart M.\ Stach$^{1}$,$\!$\thanks{E-mail: stuart.m.stach@durham.ac.uk}
U.\ Dudzevi\v{c}i\={u}t\.{e},$^{\! 1}$
Ian Smail,$^{\! 1}$
A.\,M.\ Swinbank,$^{\! 1}$
J.\,E.\ Geach,$^{\! 2}$
\newauthor
J.\,M.\ Simpson,$^{\! 3}$
Fang~Xia An,$^{\! 4,1}$
Omar Almaini,$^{\! 5}$
Vinodiran Arumugam,$^{\! 6,7}$
A.\,W.\ Blain,$^{\! 8}$
\newauthor
S.\,C.\ Chapman,$^{\! 9}$
Chian-Chou Chen,$^{\! 6}$
C.\,J.\ Conselice,$^{\! 10}$
E.\,A.\ Cooke,$^{\! 1}$
K.\,E.\,K.\ Coppin,$^{\! 2}$
\newauthor
E.\,da Cunha,$^{\! 11}$
J.\,S.\ Dunlop,$^{\! 7}$
Duncan Farrah,$^{\! 12,13}$
B.\, Gullberg,$^{\! 1}$
J.\,A.\ Hodge,$^{\! 14}$
\newauthor
R.\,J.\ Ivison,$^{\! 6,7}$
Dale D.\ Kocevski,$^{\! 15}$
M.\,J.\ Micha\l{}owski,$^{\! 16}$
Takamitsu Miyaji,$^{\! 17}$
Douglas Scott,$^{\! 18}$
\newauthor
A.\,P.\ Thomson,$^{\! 19}$
J.\,L.\ Wardlow,$^{\! 20}$
Axel Weiss,$^{\! 21}$
P.\ van der Werf$^{14}$
\\
$^{1}$Centre for Extragalactic Astronomy, Department of Physics, Durham University, Durham, DH1 3LE, UK\\
$^{2}$Centre for Astrophysics Research, School of Physics, Astronomy and Mathematics, University of Hertfordshire, Hatfield AL10 9AB, UK\\
$^{3}$Academia Sinica Institute of Astronomy and Astrophysics, No. 1, Sec.\ 4, Roosevelt Rd., Taipei 10617, Taiwan\\
$^{4}$Purple Mountain Observatory, China Academy of Sciences, 2 West Beijing Road, Nanjing 210008, China\\
$^{5}$School of Physics and Astronomy, University of Nottingham, University Park, Nottingham, NG7 2RD, UK\\
$^{6}$European Southern Observatory, Karl Schwarzschild Strasse 2, Garching, Germany\\
$^{7}$Institute for Astronomy, University of Edinburgh, Royal Observatory, Blackford Hill, Edinburgh EH9 3HJ, UK\\
$^{8}$Department of Physics and Astronomy, University of Leicester,University Road, Leicester LE1 7RH, UK\\
$^{9}$Department of Physics and Atmospheric Science, Dalhousie University Halifax, NS B3H 3J5, Canada\\
$^{10}$School of Physics and Astronomy, University of Nottingham, University Park, Nottingham, NG7 2RD, UK\\
$^{11}$Research School of Astronomy and Astrophysics, The Australian National University, Canberra ACT 2611, Australia\\
$^{12}$Department of Physics and Astronomy, University of Hawaii, 2505 Correa Road, Honolulu, HI 96822, USA\\
$^{13}$Institute for Astronomy, 2680 Woodlawn Drive, University of Hawaii, Honolulu, HI 96822, USA\\
$^{14}$Leiden Observatory, Leiden University, P.O.\ box 9513, NL-2300 RA Leiden, The Netherlands\\
$^{15}$Department of Physics and Astronomy, Colby College, Waterville, ME 04961, USA\\
$^{16}$Astronomical Observatory Institute, Faculty of Physics, Adam Mickiewicz University, ul. S\l{}oneczna 36, 60-286 Pozna\'n, Poland\\
$^{17}$Instututo de Astronom\'ia sede Ensenada, Universidad Nacional Aut\'onoma de M\'exico, Ensenada, 22860, M\'exico\\
$^{18}$Department of Physics and Astronomy, University of British Columbia, 6224 Agricultural Road, Vancouver, BC V6T 1Z1, Canada\\
$^{19}$The University of Manchester, Oxford Road, Manchester, M13 9PL, UK\\
$^{20}$Department of Physics, Lancaster University, Lancaster, LA1 4YB, UK\\
$^{21}$Max-Planck-Institut f\"{u}r Radioastronomie, Auf dem H\"{u}gel 69 D-53121 Bonn, Germany
}

\begin{document}
\label{firstpage}
\pagerange{\pageref{firstpage}--\pageref{lastpage}}
\maketitle

\begin{abstract}
We present the catalogue and properties of sources in AS2UDS, an 870-$\mu$m continuum survey with the Atacama Large Millimetre/sub-millimetre Array (ALMA) of 716 single-dish sub-millimetre sources detected in the UKIDSS/UDS field by the SCUBA-2 Cosmology Legacy Survey. In our sensitive ALMA follow-up observations we detect 708 sub-millimetre galaxies (SMGs) at  $>$\,4.3$\sigma$ significance across the $\sim$\,1-degree diameter field. We combine our precise ALMA positions with the extensive multi-wavelength coverage in the UDS field to fit the spectral energy distributions of our SMGs to derive a median redshift of $z_{\rm phot}=$\,2.61$\pm$0.09. This large sample reveals a statistically significant trend of increasing sub-millimetre flux with redshift suggestive of galaxy downsizing. 101 ALMA maps do not show a $>$\,4.3$\sigma$ SMG, but we demonstrate from stacking {\it Herschel} SPIRE observations at these positions,  that the vast majority of these blank maps correspond to real single-dish sub-millimetre sources.  We further show that these blank maps contain an excess of galaxies at $z_{\rm phot}=$\,1.5--4 compared to random fields, similar to the redshift range of the ALMA-detected SMGs. In addition, we combine X-ray and mid-infrared active galaxy nuclei activity (AGN) indicators to yield a likely range for the AGN fraction of 8--28\,\% in our sample.   Finally, we compare the redshifts of this population of high-redshift, strongly star-forming galaxies with the inferred formation redshifts of massive, passive galaxies being found out to $z\sim$\,2, finding reasonable agreement -- in support of an evolutionary connection between these two classes of massive galaxy.  
\end{abstract}



%
%
%
\section{Introduction} \label{sec:intro}

Over twenty years ago the first, deep, sub-millimetre wavelength surveys taken at the James Clerk Maxwell Telescope (JCMT) uncovered a population of sub-millimetre bright galaxies \citep[SMGs -- e.g.][]{smail1997deep,hughes1998high,barger1998submillimetre}, which were interpreted as showing some of highest rates of star formation observed in galaxies across the whole history of the Universe. Their sub-millimetre emission originates from the reprocessed ultra-violet starlight that has been absorbed by dust and re-emitted in the restframe far-infrared. This population of highly obscured galaxies are most easily selected at sub-millimetre wavelengths and so are termed 'sub-millimetre galaxies' (SMGs).


The selection of these star-forming galaxies at  sub-millimetre wavelengths has both advantages and disadvantages. A major advantage is the strongly negative $K$-correction at sub-millimetre wavelengths arising from the slope of the Rayleigh-Jeans tail of their far-infrared/sub-millimetre spectral energy distributions (SED). As a result of this negative $K$-correction, a flux limited sub-millimetre survey provides a uniform selection in terms of far-infrared luminosity (at a {\it fixed} dust temperature) for sources  across a redshift range of $z=$\,1--6 \citep[figure 4:][]{blain2002submillimeter}. Thus sub-millimetre observations are a very effective means to survey for the most strongly star-forming galaxies in the high redshift Universe. However, a major disadvantage of current single dish observatories operating at sub-millimetre wavelengths is their modest angular resolution, 15--30\arcsec\ FWHM, which is too coarse to allow the counterpart to the sub-millimetre emission to be easily identified at shorter wavelengths, as several candidate galaxies can be encompassed by the single-dish beam. Hence early attempts to pinpoint the location of SMGs to sub-arcsecond resolutions exploited the FIR--radio correlation \citep[e.g.][]{ivison1998hyperluminous,barger2000mapping,ivison2002deep,chapman2005redshift} to match the sub-millimetre sources to their radio bright counterparts. The limitation of such radio identifications is that the radio waveband does not benefit from a strong negative $K$-correction, so there is a bias against identifying the highest redshift ($z>$\,2.5--3) SMGs in the radio images. The difficulties with reliably identifying sub-millimetre source counterparts contributed in part to the slow advance in our understanding of these galaxies in the years following their discovery.

Not withstanding the challenges described above, the first large-scale spectroscopic redshift surveys of radio-identified SMGs \citep{chapman2005redshift}, and later sub/millimetre interferometrically-selected samples \citep{smolvcic2012millimeter,danielson2017alma}, found that these galaxies are typically located at redshifts of $z\sim$\,2.5. At these redshifts, the sub-millimetre flux of the sources corresponds to far-infrared luminosities $>$\,10$^{12}$--10$^{13}$\,L$_{\odot}$, i.e.\ Ultra-Luminous InfraRed Galaxies (ULIRGs). However, SMGs have volume densities three orders of magnitude greater than comparably luminous local ULIRGs \citep{chapman2005redshift}. Such high infrared luminosities indicate star-formation rates (SFR) of the order 100--1000\,M$_{\odot}$\,yr$^{-1}$ \citep{magnelli2012herschel,swinbank2014alma}, a star-formation rate high enough that within a few dynamical times (a hundred million years) the SMG could form the stellar mass of a massive galaxy $M_{\ast} \gtrsim $\,10$^{11}$\,M$_{\odot}$. Indeed, constraints on the stellar masses of SMGs have found $M_{\ast}\sim $\,10$^{11}$--10$^{12}$\,M$_{\odot}$ \citep{borys2005relationship,hainline2011stellar,michalowski2014determining} making SMGs some of the most massive galaxies at $z\sim$\,2. The space density of these sources and their prodigious star-formation rates means that SMGs contribute $\sim$\,20\% of the Universal star-formation density between $z=$\,1--4 \citep{casey2013characterization,swinbank2014alma}. Being both massive and strongly star-forming galaxies in the early Universe, SMGs have been proposed as the progenitors of massive local spheroidal galaxies  \citep[e.g.][]{genzel2003spatially,blain2004accurate,cimatti2008gmass,simpson2014alma,toft2014submillimeter,koprowski2014reassessment,simpson2017scuba}, potentially following  an evolutionary path where, following their ultra-luminous infrared phase, the SMG descendants would display  both star-formation and obscured AGN activity, and then appears as a quasi-stellar object (QSO), until the system completely exhausts its supply of gas \citep{coppin2008testing,simpson2012evolutionary}.

The major advance in studies of SMGs came with the development of sensitive sub-millimetre interferometers: initially the Sub-Millimeter Array (SMA) \citep{younger2008clarifying,wang2010sma,smolvcic2012millimeter} and more recently  the Atacama Large Millimetre/sub-millimeter Array (ALMA) \citep{hodge2013alma,simpson2015scuba,hatsukade2016sxdf,walter2016alma,franco2018goods,hatsukade2018alma,cowie2018submillimeter}. Interferometric observations, in particular with ALMA, allow us to observe SMGs in the sub-millimetre at spatial resolutions more than an order of magnitude finer than achievable in single-dish surveys and free from confusion -- enabling detections of sources to flux densities more than an order of magnitude fainter than the single-dish limits. 

Deep, blank-field surveys utilising these interferometers have successfully recovered faint, serendipitously detected sources across arcmin$^{2}$ regions such as in the {\it Hubble} Ultra-Deep Field and GOODS-South \citep{aravena2016alma,walter2016alma,dunlop2016deep,franco2018goods,hatsukade2018alma}. These are effective surveys for detecting the fainter examples of the SMG population free from the potential biases from clustering of sources around bright detections. However, the modest field of view of interferometers means that such surveys can only cover small areas and as a result have so far yielded relatively few (10's) of detected sources, with only very few of the brightest examples having $S_{870}\gg$\,1\,mJy. To obtain statistically robust samples of the brighter SMGs ($S_{870}\gs$\,1--10\,mJy), whose properties may be the most distinct from 'normal' star-forming galaxies, we require a hybrid approach -- where we exploit the fast mapping speed of single-dish telescopes to identify numbers of these relatively rare sources over the large fields needed to yield large samples -- combined with interferometric observations in the same sub-millimetre waveband to allow us to precisely locate the counterparts to the single-dish sources. We first employed this dual-survey approach with the  ALMA LABOCA Extended \emph{Chandra} Deep Field South survey (ALESS) \citep{hodge2013alma,karim2013alma}. This was an ALMA Cycle 0 survey of the 122 sub-millimetre sources detected in the LABOCA/APEX single-dish survey of the Extended \textit{Chandra} Deep Field South \citep[LESS:][]{weiss2009large} and yielded detections of 126 single-dish sources with deboosted 870\,$\mu$m fluxes $S_{870}>$\,3.6\,mJy. This survey suggested that some previous single-dish detections were in reality multiple galaxies blended by the coarse resolution of the single-dish telescope \citep{karim2013alma} and previous multiwavelength methods of identification of SMGs, were failing to correctly locate the counterpart to the sub-millimetre emission almost half of the time \citep{hodge2013alma, simpson2015scuba,simpson2015scuba2}. 

These initial ALMA studies of flux-limited samples have begun to illuminate the range of characteristics of bright sub-millimetre galaxies, free from the selection biases which influenced earlier radio and mid-infrared based studies. In particular they have highlighted the $\sim$\,10--20\% of the population which are effectively undetectable in even the deepest optical/near-infrared \citep{simpson2014alma}, which  represent either the highest redshift, the least massive or the most obscured examples of this population.  However, the first ALMA surveys lacked the sample size to identify statistically significant subsets of the rarest classes of SMGs. For example there are only two $z\sim$\,4.4 [C{\sc ii}]-selected sources in the ALESS survey \citep{swinbank2012alma,gullberg2018dust}, which provide an insight into the properties of the more distant examples of sub-millimetre galaxies. Similarly, ALESS yielded  just ten X-ray detected AGN--SMG systems, which can be used to probe the co-evolution of super-massive black holes in strongly star-forming galaxies \citep{wang2013alma}.  To improve the statistical strength of the conclusions about these rarer subclasses of sub-millimetre galaxies, larger surveys are needed in extragalactic survey fields with the deepest supporting data necessary to detect the faintest examples of this population.

Driven by this need, we have just completed a larger study, nearly an order-of-magnitude larger than ALESS, which exploits the wide-field sub-millimetre mapping of key extragalactic survey fields undertaken by the SCUBA-2 Cosmology Legacy Survey \citep[S2CLS:][]{geach2017scuba}.  We focus in this project on the S2CLS 850-$\mu$m map of the $\sim$\,1\,degree diameter UKIDSS Ultra Deep Survey (UDS) Field, which was the largest, uniform area mapped by S2CLS.  The S2CLS UDS map has a median sensitivity of $\sigma_{850}=$\,0.9\,mJy over an area of 0.96\,degrees$^2$, with 716 sources catalogued above a 4-$\sigma$ detection limit (corresponding to a 2\,\% false positive rate) of $S_{850}\sim$\,3.5\,mJy.  We began our investigation of this sample with a pilot ALMA study of a  subset of thirty bright SCUBA-2 detected sources in Cycle 1 \citep{simpson2015scuba,simpson2015scuba2,simpson2017scuba}.  We then expanded the study during Cycles 3, 4, and 5 to complete the ALMA 870-$\mu$m observations of all 716 $>$\,4$\sigma$ sources.  This yields AS2UDS -- the ALMA SCUBA-2 UDS survey -- the largest, homogeneously selected, sample of SMGs to date with 708 detections, a five-fold increase over the previous largest similarly robust sample. The first results from this survey have already been presented: number counts and rates of multiplicity \citep{stach2018alma}, the serendipitous detection of high redshift [C{\sc ii}] emitters \citep{cooke2018alma}, and the use of this survey as a training set for machine learning algorithms to identify the multiwavelength counterparts to single-dish submillimetre sources \citep{an2018machine}.   We present a full analysis of the multiwavelength properties of this sample in Dudzevi\v{c}i\={u}t\.{e} et al.\ (in prep.) and in Gullberg et al.\ (in prep.) we discuss the information available on the sizes and morphologies of the  dust continuum in these sources from our highest resolution ALMA observations. 

In this paper we present the final catalogue for the AS2UDS survey\footnote{Catalogue can be found at http://astro.dur.ac.uk/AS2UDS}. In our analysis we compare results from our new large sample to previous studies.  To simplify these comparisons we have limited them in general to flux-limited samples from: 1.\ larger unbiased blank-field surveys at 850\,$\mu$m \citep[as there is evidence of differences compared to populations selected in the far-infrared and millimetre, e.g.][]{smolvcic2012millimeter,koprowski2014reassessment,scudder2016multiplicity,ikarashi2017very}; 2.\ with deep ($<$\,1\,mJy rms) interferometric identifications in the same waveband as any initial single-dish selection, if appropriate \citep[c.f.][]{barger2014there,umehata2014aztec,hill2018high}; 3.\ and which are not explicitly lensed, owing to the potential selection effects and variable flux limits as well as uncertainties from cluster lenses and especially galaxy-scale lensed samples \citep[e.g.][]{weiss2013alma,fujimoto2015alma,arancibia2018alma}. Thus most of our comparisons are made to the ALESS survey \citep{hodge2013alma}, SuperGOODS \citep{cowie2018submillimeter} and the various ALMA surveys in GOODS-S \citep{walter2016alma,dunlop2016deep,franco2018goods,hatsukade2018alma}.

In \S\ref{sec:datareduction} we describe the target selection for our ALMA survey and data reduction across the different ALMA Cycles and the wealth of multi-wavelength archival data available in this field, most notably from the UKIDSS UDS DR11 catalogue (O.\ Almaini et al.\ in prep.). \S\ref{sec:source} describes the source detection algorithm, the simulated maps used for estimating completeness and flux boosting derivations. In \S\ref{sec:results} we present the first results from our \textsc{magphys} SED fitting (Dudzevi\v{c}i\={u}t\.{e} et al.\ in prep.): the photometric redshift distribution of our sample and comparisons with previous surveys. In addition we present the selection of active galactic nuclei (AGNs) from our catalogue through archival X-ray observations of the field and IRAC colour-colour selection. \S\ref{sec:conclusions} presents our main conclusions. We assume a cosmology with $\Omega_{\rm m}=$\,0.3, $\Omega_{\Lambda}=$\,0.7, and H$_{0}=$\,70\,km\,s$^{-1}$\,Mpc$^{-1}$. All magnitudes are in the AB system and errors are calculated from bootstrap analysis unless otherwise stated.

\section{Observations and Data Reduction} \label{sec:datareduction}

\subsection{Sample Selection}

\begin{figure*}
\includegraphics[width=2\columnwidth]{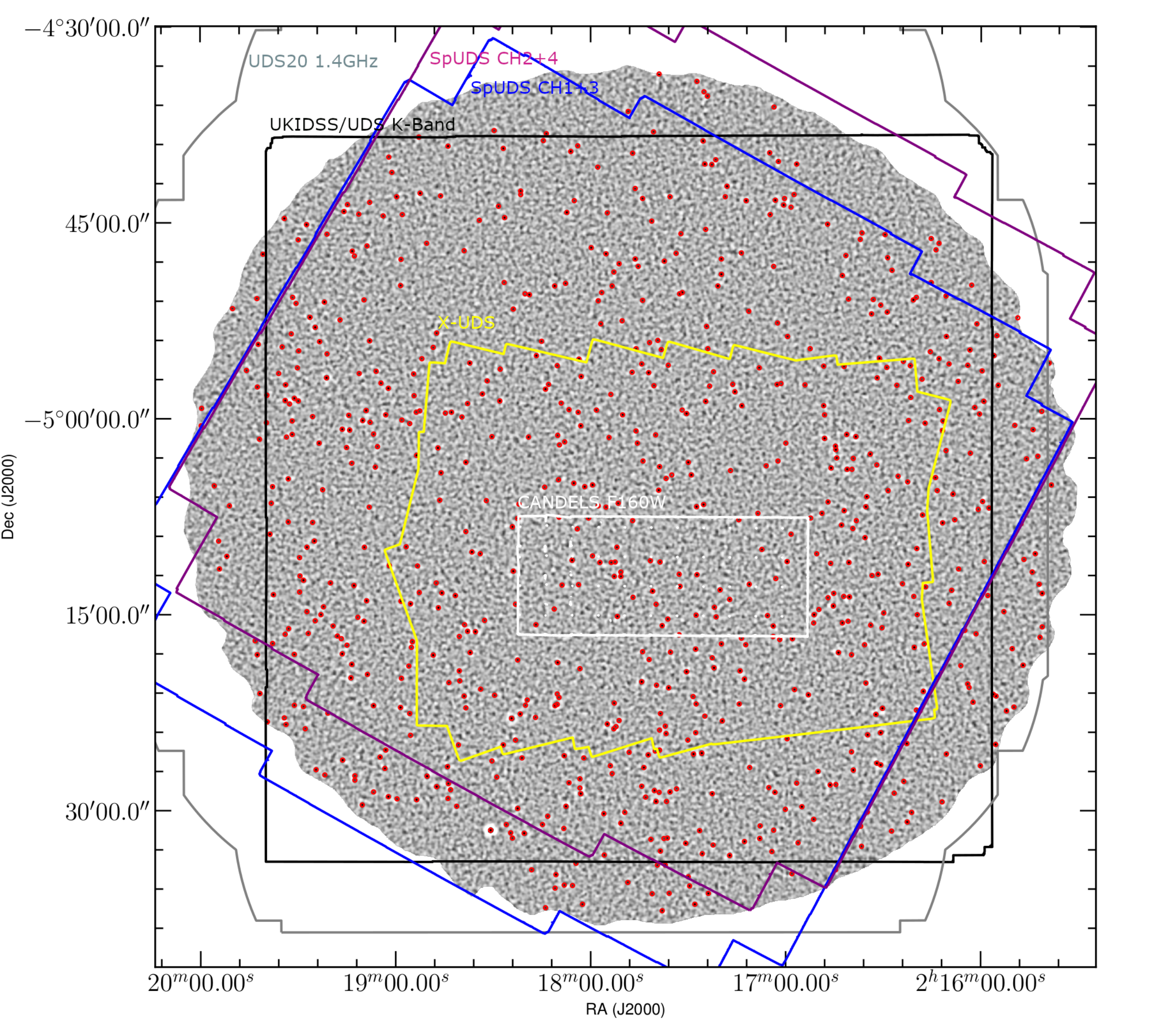}
\caption{A sample of the multi-wavelength coverage of the AS2UDS sample which motivates this extra-galactic field being chosen for high resolution ALMA follow-up. The background shows the S2CLS 850\,$\mu$m UDS map from which the parent sample is extracted \citep{geach2017scuba}. The red circles are the primary beams for AS2UDS targetting the 4\,$\sigma$ detections in the S2CLS map. The black outline shows the $K$-band coverage that forms the footprint for the UKIDSS UDS catalogue (O.\ Almaini et al.\ in prep.). The \emph{Spitzer}/IRAC CH1/3 and CH2/4 coverage is shown in blue/purple respectively, the \emph{HST} CANDELS F160W in white, VLA 1.4\,GHz in grey, and the X-UDS \emph{Chandra} coverage in yellow \citep{kocevski2018x}.}
\label{fig:coverage}
\end{figure*}

The ALMA--SCUBA-2 Ultra Deep Survey, hereafter AS2UDS, is a high-resolution, sub-millimetre interferometric follow-up survey of the SCUBA-2 850\,$\mu$m sources selected from the S2CLS map of the UDS field (Fig\,\ref{fig:coverage}). The parent single-dish survey covers an area of 0.96\,deg$^{2}$, with noise levels below 1.3\,mJy and a median depth of $\sigma_{850}=$\,0.88\,mJy\,beam$^{-1}$ with 80\% of sources identified in regions of the map with $\sigma_{850}=$\,0.86--1.02\,mJy\,beam$^{-1}$ \citep{geach2017scuba}. Across four ALMA cycles (1, 3, 4, and 5) we observed all 716 $>4\sigma$ sources from the SCUBA-2 S2CLS map (corresponding to observed flux densities $S_{850}\geq $\,3.4\,mJy) (see: Figure\,\ref{fig:coverage}). 

%
%

In Cycle 1 (Project ID: 2012.1.00090.S), 30 of the brightest sources from an early version of the SCUBA-2 UDS map (data taken before 2013 February) were observed in ALMA Band 7 \citep{simpson2015scuba,simpson2015scuba2,simpson2017scuba}. This early version of the SCUBA-2 map had a depth of only $\sigma_{850}\sim$\,2.0\,mJy\,$^{-1}$ and subsequent integration time in the S2CLS survey scattered three of these sources below our final sample selection criteria ($>$\,4$\sigma$), leaving 27 of these original single-dish detected sources in our final sample. The remaining 689 single-dish sources in the final S2CLS catalogue were observed across ALMA cycles 3 and 4 (Project ID: 2015.1.01528.S and 2016.1.00434.S, respectively). To cross calibrate the data, a fraction of these sources were observed twice, once in Cycle 3 \textit{and} 4. In addition, in Cycle 5 ten of the brightest SCUBA-2 sources which returned `blank' maps from the Cycle 3 and 4 ALMA observations were re-observed at greater depth (Project ID: 2017.1.01492.S), these maps will be discussed further in \S\ref{sec:blanks}.
\vspace{1mm}

\subsection{Data Reduction} \label{sec:data}
Our ALMA targets were observed in Band 7 (central frequency 344\,GHz\, $\sim$\,870\,$\mu$m). At this frequency the FWHM of the ALMA primary beam (17$\farcs$3) covers the FWHM of the SCUBA-2 beam (14$\farcs$7). Cycle 1 observations were carried out on 2013 November 1 \citep{simpson2015scuba}, Cycle 3 between 2016 July 23 and August 11, Cycle 4 between 2016 November 9 and 17 and 2017 May 6, and Cycle 5 on 2018 August 24. 

These ALMA Band 7 continuum observations were 150\,second integrations in Cycle 1 using 26 dishes, 40\,second integrations in Cycle 3 and 4 (with 45--50 dishes), and 285\,second integrations using 44 dishes in Cycle 5, with the 7.5\,GHz bandwidth of the four spectral windows centred at 344.00\,GHz. The array configurations for Cycle 1 observations yielded a median synthesised beam size of 0$\farcs$35$\times$0$\farcs$25. With Cycle 3 and 4 observations we aimed to match this resolution, however our Cycle 3 observations were taken with a more extended array configurations resulting in median synthesised beam sizes of 0$\farcs$19$\times$0$\farcs$18 (`natural' weighting). Our Cycle 5 observations were intended to test whether the lack of detected counterparts in the ALMA observations of ten SCUBA-2 sources observed in previous cycles was not a result of flux being resolved out, and therefore were taken with a median synthesised beam size of 0$\farcs$81$\times$0$\farcs$54.

Phase centres for each observation were set to the SCUBA-2 positions of the S2CLS catalogue at the time of ALMA proposal submissions. For the 689 SCUBA-2 sources followed up in Cycle 3, 4, and 5 this results in phase centres coincident to the SCUBA-2 positions from the S2CLS DR1 sub-millimetre source catalogue \citep{geach2017scuba},  however for Cycle 1 the 27 SCUBA-2 source positions were not corrected for a systematic astrometric offset in the interim map and an offset exists between the final S2CLS DR1 source position and ALMA phase centres, with a median offset of 3.2$^{+0.1}_{-0.6}$\,arcsec. The ALMA primary beam size is large enough that this offset still results in the majority ($\sim$\,95\%) of the SCUBA-2 beam falling within the primary beam of in our 30 Cycle 1 observations and, as discussed below, these are observations which were targeting the brighter SCUBA-2 sources and thus we expect detections closer to the offset phase centres and thus still well within the ALMA primary beams.

The \textsc{Common Astronomy Software Application} \citep[\textsc{CASA}][]{mcmullin2007casa} v4.1, v4.5.3, v4.7.2, v5.3.0 were used to calibrate the datasets from Cycle 1, 3, 4, and 5 respectively using the standard ALMA calibration scripts. For the bandpass and phase calibration observations of J0006$-$0623, J0423$-$0120, J0238+1636, J0241$-$0815, and J0006$-$0623 were used across the four cycles and for the absolute flux scaling: J0217$-$0820 and J0238+1636. 

For the imaging \textsc{CASA} v 4.7.2 was used, with the \textsc{concat} task to combine the visibilities of the 125 maps that were observed in both Cycle 3 and Cycle 4. To create the continuum maps we used the \textsc{clean} task in multi-frequency synthesis mode. Previous studies on the sizes of sub-millimetre galaxies suggests their sub-millimetre intrinsic sizes are 0$\farcs$3$\pm$0$\farcs$1 \citep[e.g.][]{simpson2015scuba2,hodge2016kiloparsec,ikarashi2017very} therefore we expect the majority of our detections to be marginally resolved. To ensure we are not biased against selecting extended sources, by remaining sensitive to extended flux from our SMGs, we employed a 0$\farcs$5 FWHM Gaussian taper in the \textit{uv}-plane using the \emph{uvtaper} parameter in \textsc{clean} for the Cycle 1,3, and 4 data. This effectively down-weights the visibilities from the longer baselines thus enhancing the sensitivity to extended emission at the expense of overall reduced sensitivity in the maps. Combined with natural weighting this resulted in `detection' maps with median synthesized beam sizes of 0$\farcs$73\,$\times$\,0$\farcs$59 for Cycle 1, 0$\farcs$56\,$\times$\,0$\farcs$50 for Cycle 3 and 0$\farcs$58\,$\times$\,0$\farcs$55 for Cycle 4 (the untapered Cycle 5 maps has beam sizes 0$\farcs$79\,$\times$\,0$\farcs$52).

\begin{figure}
\includegraphics[width=\columnwidth]{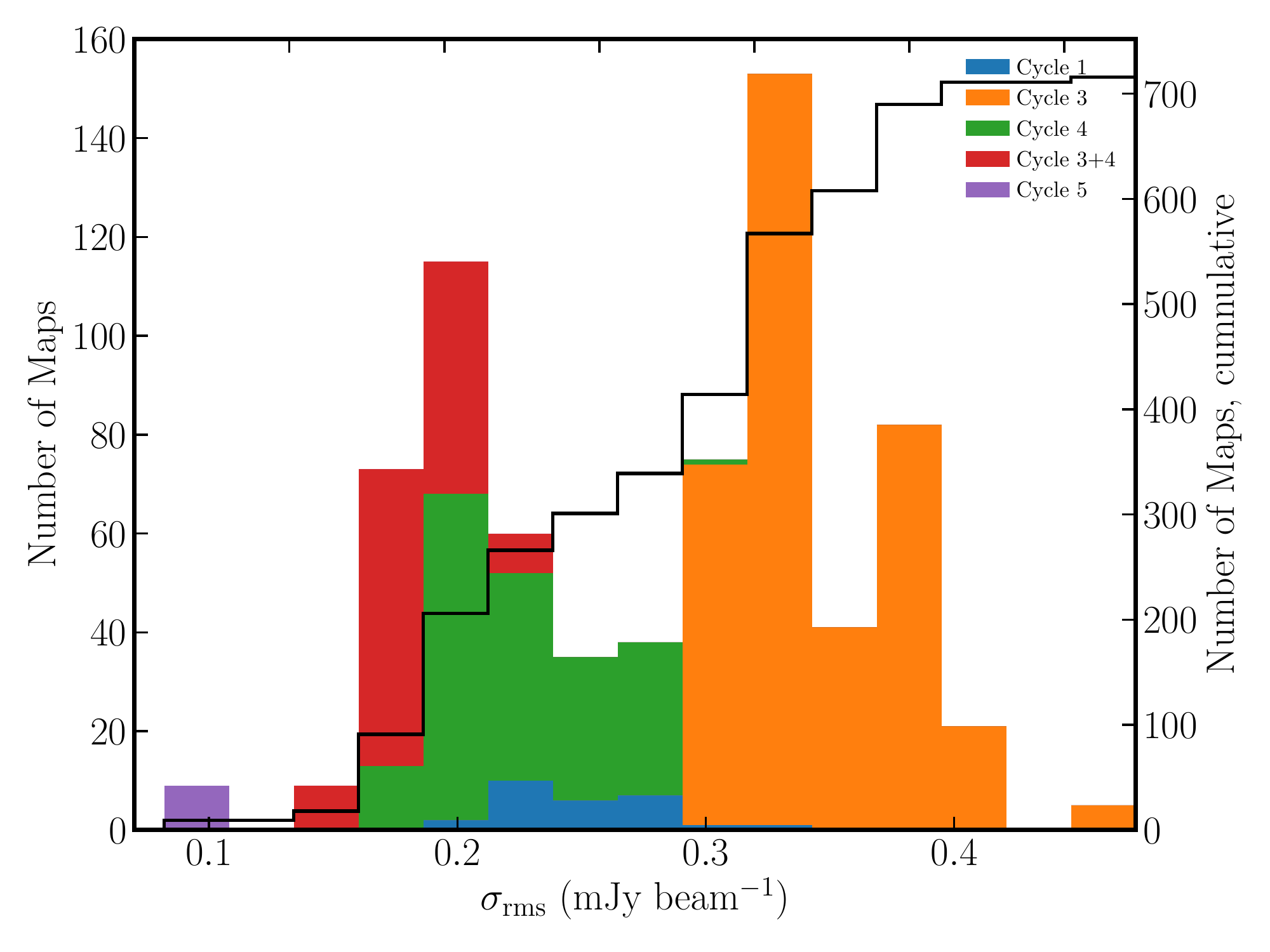}
\caption{The distribution of the rms noise estimates from the 0$\farcs$5 $uv$-tapered detection maps for the different cycles used in AS2UDS. Adopting similar reconstructed beam sizes across all our maps results in the final rms noise varying inversely with their original observed resolution, with the exception of the purposefully much deeper `blank' field repeat observations in Cycle 5. Deviations in rms across the same cycle are a result of different observing conditions. Overlaid is the cumulative distribution of the rms which shows after tapering the median rms of the full survey is 0.3$^{+0.1}_{-0.2}$\,mJy\,beam$^{-1}$.}
\label{fig:rms}
\end{figure}

For the ALMA map cleaning, the first step was creating `dirty' maps, 512$\times$512 pixels in size with a pixel scale of 0$\farcs$06 (30$\farcs$7 square), by combining the spectral windows with no cleaning cycles applied. We calculated the average root mean square (rms) flux density for each of these `dirty' maps by first removing pixels associated with potential sources in the maps by applying an iterative sigma-clipping routine to the map, and then finding the root mean square value of the unclipped values ($\sigma$). Each map was then initially cleaned to 3\,$\sigma$, and islands of pixels with flux densities above 4\,$\sigma$ were convolved with the reconstructed beam to create cleaning masks. If no islands were detected then the initially-cleaned map was taken as the final map however if any sources were detected then the clean mask was applied and sources cleaned to 1.5\,$\sigma$. This resulted in final cleaned `detection' maps which have mean depths of $\sigma_{\rm 870}=$\,0.25\,mJy\,beam$^{-1}$ for Cycle 1, $\sigma_{870}=$\,0.34\,mJy\,beam$^{-1}$ for Cycle 3, $\sigma_{\rm 870}=$\,0.23\,mJy\,beam$^{-1}$ for Cycle 4, and $\sigma_{870}=$\,0.085\,mJy\,beam$^{-1}$ for Cycle 5. Figure~\ref{fig:rms} shows the histogram of the depths for each of the 716 images with the median depth across the survey of 0.3$^{+0.1}_{-0.2}$\,mJy\,beam$^{-1}$, which is $\sim$\,3$\times$ deeper than the original single-dish SCUBA-2 survey but with a beam $\sim$\,600$\times$ smaller in area -- allowing us to precisely locate the source(s) of the sub-millimetre emission seen by SCUBA-2.

%
%

\subsection{Additional Multiwavelength Observations} \label{sec:multi}

In this paper we present some of the basic properties of our sub-millimetre catalogue, using redshifts derived from spectral energy distribution (SED) fitting with  \textsc{Multi-wavelength Analysis of Galaxy Physical Properties} (\textsc{magphys}) \citep{da2008simple}. This analysis exploits the wealth of deep, multi-wavelength observations available in this field (e.g.\ Figure\,\ref{fig:coverage}). In this section we describe the multi-wavelength observations used in the \textsc{magphys} analysis, however the full description of the \textsc{magphys} SED fitting and the  resulting constraints on the source properties are given in  Dudzevi\v{c}i\={u}t\.{e} et al.\ (in prep.).

The basis of our multi-wavelength analysis is taken from the UKIRT Infrared Deep Sky Survey (UKIDSS) Ultra Deep Survey data release 11 catalogue (O.\ Almaini et al.\ in prep.). This survey contains 296,007 $K$-band detected sources, extracted with 2\arcsec diameter apertures corrected to total galaxy magnitudes, detected with \textsc{sextractor} with photometry retrieved from $J$ and $H$ maps using \textsc{sextractor} dual-image mode. These $J, H,$ and $K$-band images were observed through a mosaic of four observations, covering a total survey area of 0.77\,deg$^{2}$, with the Wide-Field Camera at UKIRT \citep{casali2007ukirt}, which covers 643 of the 716 pointings of AS2UDS, or $\sim$\,90\%. The DR11 maps achieve a 2\arcsec aperture 3-$\sigma$ median depth of $J=$\,26.2, $H=$\,25.7, and $K=$\,25.9\,mag making this one of the deepest near-infrared surveys on degree scales. The DR11 UDS $K$-band selected catalogue has been matched with a number of other surveys to broaden the wavelength coverage, with the matching and photometry measurements described in Hartley et al.\ (in prep.). We use a 0$\farcs$6 matching radius to cross-match the DR11 UDS catalogue to our AS2UDS sub-millimetre galaxy catalogue. This matching radius provides a low false match ratio ($\sim$\,3.5\,\%) when matching the two catalogues \citep[see:][]{an2018machine}.

Complementary optical data comes from the Subaru/\textit{XMM-Newton} deep survey (SXDS) \citep{furusawa2008subaru}, this is a survey with $B$, $V$, $R_{c}$, $i'$ and $z'$-band magnitudes with 2\arcsec aperture 3-$\sigma$ depths of $B=$\,28.4, $V=$\,27.8, $R_{c}=$\,27.7, $i'=$\,27.7 and $z'=$\,26.6\,mag. Additionally there is $Y$-band data with a 3-$\sigma$ depth of 25.3\,mag supplied from the Visible and Infrared Survey Telescope for Astronomy (VISTA) Deep Extra-galactic Observations (VIDEO) survey \citep{jarvis2012vista}.

We include observations in the near-infrared from the \textit{Spitzer} UKIDSS Ultra Deep Survey (SpUDS; PI: J.\ Dunlop), a $\sim$\,1\,deg$^{2}$ IRAC (at 3.6, 4.5, 5.8, and 8.0\,$\mu$m -- corresponding to channels: Ch1, 2, 3, and 4 respectively) survey of the UDS field with 3-$\sigma$ limiting depths of 23.5, 23.3, 22.3, and 22.4\,mag in Ch1--4 respectively. The astrometry of the IRAC images was corrected by stacking the IRAC images on the DR11 UDS $K$-band locations and corrections of +0$\farcs$00 R.A.\ and +0$\farcs$15 Dec.\ were applied to the Ch1 image, +0$\farcs$075 R.A.\ and +0$\farcs$12 Dec.\ to Ch2, +0$\farcs$075 R.A.\ and +0$\farcs$0 Dec.\ to Ch3, and +$\farcs$6 R.A.\ $-$0$\farcs$075 Dec.\ to Ch4. At each of the AS2UDS galaxy locations 2$\farcs$0 aperture corrected magnitudes were measured, in order to be consistent with the other optical photometric bands. We checked the quality of our photometry by comparing our Ch1 and Ch2 aperture corrected magnitudes to those given in the UKIDSS DR11 catalogue (Ch3 and Ch4 are not supplied in the DR11 catalogue). All sources with a neighbour within 2$\farcs$5 were checked for possible contamination. Conservatively, we calculated how $K$-band aperture corrected magnitudes (or in some cases $K$-band magnitude limits) of the AS2UDS sources and the near-by sources would change if observed at the resolution of the IRAC Ch1 data. If the flux from a nearby source -- as measured in a 2$\farcs$0 aperture at the position of AS2UDS source -- resulted in contamination of over 50\,\%, then IRAC magnitudes were set to limits. This reduced the number of detection in each IRAC band by 111.

\subsubsection{Far-Infrared}
We include photometry at 100, 160, 250, 350 and 500\,$\mu$m, where available, from the \textit{Herschel} Multi-tiered Extragalactic Survey \citep[HerMES;][]{oliver2012herschel}. To correct the astrometry of the SPIRE images the same shifts of $<$\,1\farcs5, found in the AS2UDS pilot sample of \cite{simpson2017scuba}, were applied to the \textit{Herschel}/SPIRE and PACS images. This shift was derived from re-centering SPIRE stacked images using VLA radio source positions (see \S\,\ref{sec:radio}). We have confirmed that the radio astrometry aligns with ALMA to $<$\,0\farcs1 in both R.A.\ and Dec.\ and so no additional correction to the SPIRE astrometry is required.
SPIRE has comparatively low angular resolution with 18, 25, and 35\arcsec FWHM at 250, 350 and 500\,$\mu$m respectively and therefore to deblend the low resolution images we apply the technique described in \cite{swinbank2014alma}. The flux errors, and the detection limits of 5.2, 12.1, 9.2, 10.6, and 12.2\, mJy at 100, 160, 250, 350 and 500\,$\mu$m respectively are derived from simulations \citep[see:][]{swinbank2014alma}.

\subsubsection{Hubble Space Telescope}

\begin{figure*}
\includegraphics[width=2\columnwidth]{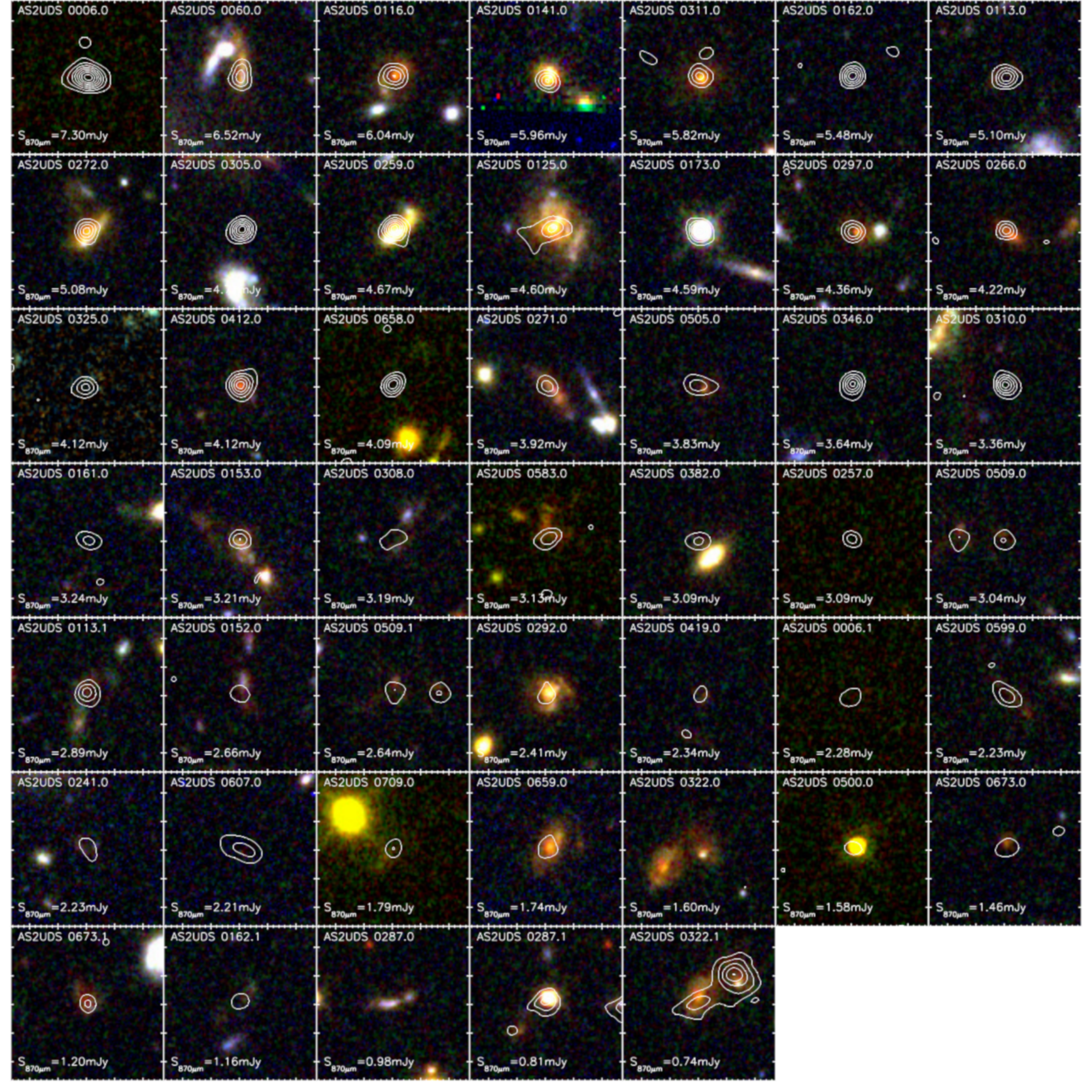}
\caption{\emph{HST} $I_{814}J_{125}H_{160}$-band colour images (5$\farcs$4 square) of the 47 ALMA SMGs in our sample that lie in the CANDELS region.  Contours are taken from the tapered ALMA maps, and denote the 870\,$\mu$m emission, and start at 3\,$\sigma$ and are incremented by 3\,$\sigma$.  The 870\,$\mu$m flux (in mJy) is given in the lower left corner.  The \emph{HST} morphologies of the ALMA SMGs display a range of morphologies, although the majority are morphologically complex. In general the sub-mm emission also appears more compact than the rest-frame optical emission \citep[e.g.][]{simpson2015scuba2}.
}
\label{fig:candels}
\end{figure*}

A 201.7\,arcmin$^{2}$ region of the UDS field (shown in Figure \ref{fig:coverage}) was covered by the \textit{Hubble Space Telescope} (\emph{HST}) in the Cosmic Assembly Near-infrared Deep Extragalactic Legacy Survey \citep[CANDELS:][]{galametz2013}. This provides a WFC3 F160W ($H_{160}$) selected catalogue of sources with a 5-$\sigma$ limiting magnitude of 27.45\,mag for a point source, with \textsc{sextractor} dual-mode source extraction in the F814W ($I_{814}$), and F125W ($J_{125}$) bands, 47 of our SMGs are covered in the CANDELS region, and we show these in Figure \ref{fig:candels}.

%
%

\subsubsection{Radio} \label{sec:radio}

Imaging at 1.4\,GHz is supplied from part of the UDS20 survey (V.\ Arumugam et al.\ in prep.) which covers $\sim$\,1.3\,deg$^{2}$, and as can be seen in Figure\,\ref{fig:coverage} this mosaic covers the near entirety of the pointings in AS2UDS (714/716 sources). The radio map averages a 1-$\sigma$ depth of 10\,$\mu$Jy\,beam$^{-1}$ with the deepest regions of the map reaching 7\,$\mu$Jy\,beam$^{-1}$ with a synthesised beam size of $\sim$\,1\farcs8 FWHM. We employ a 1\farcs6 matching radius from the AS2UDS sources to the VLA 4-$\sigma$ catalogue given by Arumuham et al.\ (in prep.) as this is the radius at which the cumulative number of VLA detections flattens, which yields a false matching rate of 1\,\%. For our 708 SMGs, 706 are covered by the UDS20 survey. Of those, 273 have a radio counterpart within 1\farcs6 (29\,\%) however this includes close pairs in the AS2UDS catalogue which match to a single radio source therefore only 264 unique radio sources are matched to the AS2UDS catalogue.

\subsubsection{X-ray}\label{sec:xray}

Deep \textit{Chandra} observations of part of the UDS field have been obtained by  the X-UDS survey \citep{kocevski2018x}. This survey covers 0.33\,deg$^{2}$ centred around the {\it HST}/CANDELS survey region (Figure~\ref{fig:coverage}). This coverage comprises a deep centre and shallower  coverage over a wider area.  In the central $\sim$\,100\,arcmin$^{2}$ of this region, the survey has an average exposure time per pixel of 600\,ks and outside of this area the survey has an exposure of 200\,ks.
In total the X-UDS catalogue has 868 X-ray point sources above a flux limit of 4.4\,$\times$\,10$^{-16}$ ergs\,s$^{-1}$\,cm$^{-2}$ in the full band (0.5--10\,keV). 

We match the X-UDS catalogue to the AS2UDS catalogue using matching radii based on the positional errors radii in the X-UDS survey (median positional error of 0$\farcs$96). Over the full X-UDS coverage we have 274 SMGs falling within the {\it Chandra} footprint, of which 21 are matched to X-ray counterparts.  Within the CANDELS {\it HST}  area, where the \textit{Chandra} coverage is deepest, we have 47 SMGs, but only two of these match to X-UDS sources.   We also perform a stacking analysis to derive average X-ray fluxes for samples of individually undetected SMGs.  This analysis uses the  X-UDS \textit{Chandra} soft (0.5--2 keV) and hard (2--8 keV) band observations and the \textsc{cstack} stacking software developed by T.\ Miyaji.  For samples of SMGs we use \textsc{cstack} to determine the mean stacked, background subtracted, count-rates and uncertainties from which fluxes were derived using the count rate to flux conversion factors given in \citet{kocevski2018x}.  These fluxes are then converted into X-ray luminosities by assuming a power-law X-ray spectrum with photon index $\Gamma=$\,1.7, consistent with the SED shape assumed for the X-ray detections in \citet{kocevski2018x}, and the photometric redshifts estimated below.

\section{Analysis} \label{sec:source}

For each ALMA map, 0$\farcs$5 diameter aperture noise levels are derived from randomly placing 50 apertures within the area of the primary beam and calculating the standard deviation of the resulting aperture fluxes. This aperture size roughly matches the $uv$-taper size, however, being slightly smaller in area than the resulting ALMA restoring beam sizes, results in 0$\farcs$5 aperture depths approximately a factor of two deeper than the noise per beams quoted above. We extract sources by initially using \textsc{sextractor} to find $>$\,2\,$\sigma$ peaks in the `detection' maps. At each of these detections a 0$\farcs$5 diameter aperture flux was measured and, from the aperture noise levels, signal-to-noise ratio (SNR) values calculated for each source.

The AS2UDS catalogue comprises the 708 sources which have 0$\farcs$5-diameter aperture SNR\,$\geq$\,4.3. This threshold, as well as the 0$\farcs$5 aperture size were chosen to maximise the number of sources detected whilst not exceeding a 2\,\% contamination rate. This is based on the ratio of `negative' sources found in the inverted ALMA maps (i.e.\ multiply pixel values by $-$1) compared to the total number of sources detected from both the inverted and regular maps as a function of the signal-to-noise threshold.

To measure the flux density of our sources, 1$\farcs$0 diameter aperture fluxes were extracted for the 708 sources from the primary beam corrected maps. A 20\,\% aperture correction is applied to the fluxes. This correction is derived from aperture fluxes taken from the flux calibrator maps imaged in the same manner as the science targets. We note that our typical sources are not expected to be well-resolved in these 0$\farcs$5-tapered maps.  We discuss the information available on the sizes and morphologies of the AS2UDS SMGs from the untapered   higher-resolution observations in Gullberg et al.\ (in prep.).

\subsection{Completeness and Flux Deboosting}

The completeness of the AS2UDS catalogue and flux deboosting factors were estimated from a suite of 60,000 simulated ALMA maps, as described in \cite{stach2018alma}. Briefly, the background noise for these maps were taken by selecting ten AS2UDS maps that sampled the $\sigma_{870}$ distribution for the complete survey. The residual maps (the observed ALMA maps minus the flux from any detected sources) for these ten maps were extracted from {\sc casa}. One simulated source was then injected into each of these noise maps, to avoid the possibility of multiple injected sources being injected into the same beam location. Following the approach in previous work \citep{karim2013alma,simpson2015scuba2} the fluxes for these sources were randomly sampled from a steeply declining power-law distribution with an index of $-$2. The FWHM sizes of the sources were randomly selected from a uniform (and hence conservative) distribution between point sources and 0$\farcs$9 FWHM extended sources. These intrinsic sources were then convolved with the ALMA synthesised beam corresponding to the original residual map the simulated source was allocated to, in a random location within the ALMA primary beam. The above described source extraction method was applied to each of the simulated maps and a successful recovery logged when a detection was found within one synthesised beam FWHM, or 0$\farcs$6 of the injected source position.

\begin{figure*}
\includegraphics[width=2\columnwidth]{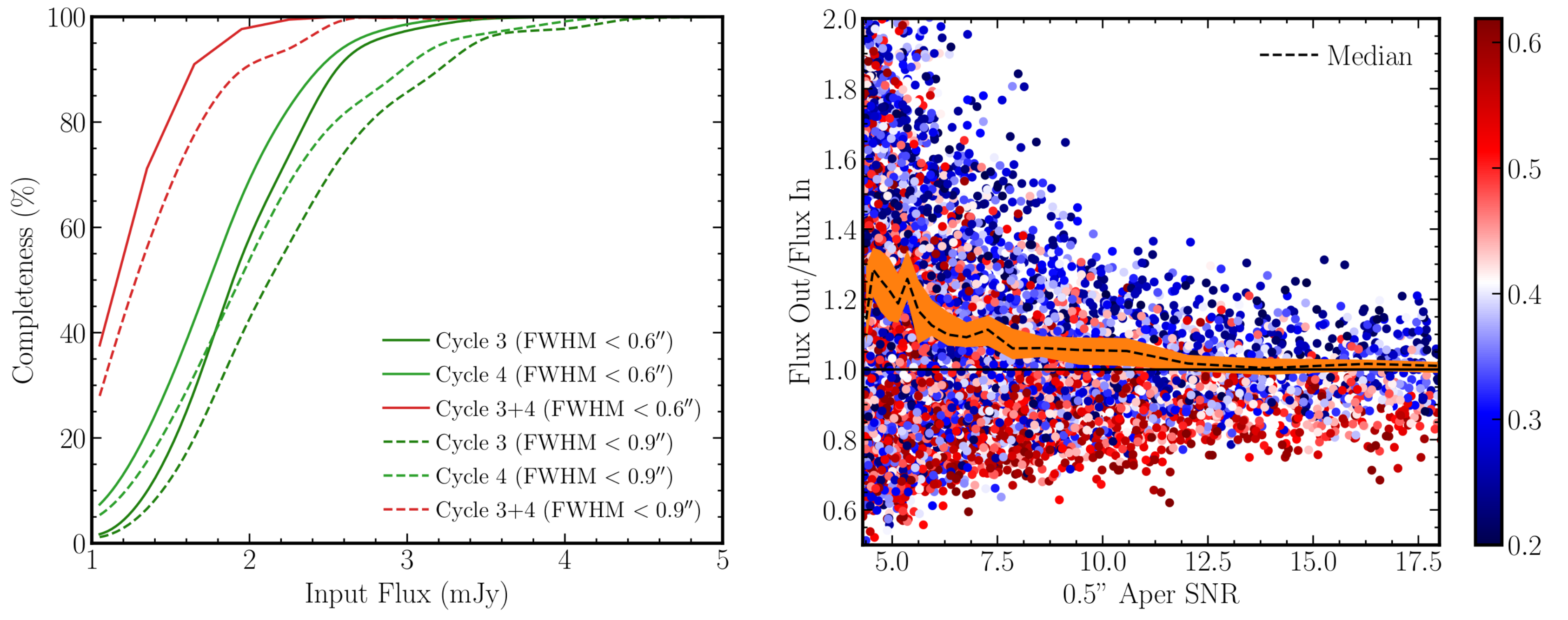}
\caption{\textit{Left:} Completeness fractions as a function of the input total flux for each simulated galaxy estimated from 60,000 simulated ALMA observations. As described in \citet{stach2018alma} our catalogue is complete for sources above $S_{850}\geq$\,4\,mJy. However, below this flux limit the completeness declines, reaching  50\,\%  at 1.3\,mJy for the ALMA maps that were observed twice. The solid lines show the completeness fractions for simulated sources with intrinsic FWHM\,$<$0$\farcs$6 whilst the dashed line shows a more conservative estimate including all simulated sources (with sizes up to 0$\farcs$9). \textit{Right:} The ratio of the recovered to injected fluxes for our simulated galaxies showing the influence of flux boosting, where the lower signal-to-noise sources have measured fluxes  boosted relatively higher by noise fluctuations. To correct for this we calculate the median boosting (dashed line) with the shaded region showing the 1\,$\sigma$ bootstrap error. Using the power-law distribution of simulated sources we find a boosting factor of 32\,\% for sources at our 4.3\,$\sigma$ threshold. The colour coding of the points is based on the intrinsic FWHM size of the simulated sources using the colour bar on the right (scale in arcseconds). As the extended sources  have systematically lower recovered fluxes this tends to reduce  the flux deboosting factors.}
\label{fig:5}
\end{figure*}

The intrinsic sizes of the injected sources has a noticeable effect on the recovered completeness fractions, particularly at lower signal-to-noise \citep[see also:][]{franco2018goods}. In Figure \ref{fig:5} we compare the completeness fractions for the recovered sources in Cycle 3, Cycle 4, and Cycle 3+4 for the complete simulated sample and the completeness fractions when cutting simulated sources with intrinsic FWHM\,$<$\,0$\farcs$6. At the highest signal-to-noise, and at the original SCUBA-2 flux limit of $S_{850}\geq$\,3.6\,mJy, AS2UDS is 98\,$\pm$\,1\,\% complete for all simulated sources with only the most extended sources suffering some  incompleteness \citep[see:][]{stach2018alma}. The difference in completeness ratios at the fainter end however becomes more pronounced with a 1.5\,mJy source having a median 52$\pm$4\% chance of detection with an intrinsic source size $<$\,0$\farcs$3, 25$\pm$2\% for sizes 0$\farcs$3$--$0$\farcs$6, and just 7$\pm$1\% for sizes $>$\,0$\farcs$6. 

%
%

Next, we estimate the effect of flux boosting. The flux boosting is a consequence of the tendency of low signal-to-noise ratio sources to have their measured fluxes preferentially increased by noise fluctuations in the maps. We estimate the magnitude of this effect by taking the median ratios of the recovered flux density for each of the detected simulated sources with the known input flux density as a function of the signal-to-noise ratio as shown in Figure~\ref{fig:5}. As with the completeness calculations, we show that the intrinsic source size distribution can have a noticeable effect on the flux deboosting. In Figure~\ref{fig:5} we show the flux recovery as a function of intrinsic source FWHM which indicates that the recovered flux of extended sources is systematically low, this therefore brings the flux deboosting values down. As previous studies suggest median sub-millimetre emission sizes of $\sim$\,0\farcs3 \citep{tacconi2006high,simpson2015scuba} we follow \cite{simpson2015scuba} and limit our simulated sources to FWHM\,$<$\,0$\farcs$6. This results in flux deboosting factors of 32\,\% for sources at our 4.3\,$\sigma$ detection threshold, reducing to 10\,\% at 7.8\,$\sigma$ and $\sim$\,0\,\% at 12\,$\sigma$. A running median is calculated from this ratio as a function of SNR in a 0$\farcs$5 diameter aperture, shown in Figure.~\ref{fig:5}, and this median value is used to correct the boosted fluxes to derive our final aperture corrected, flux deboosted, flux densities.

\subsection{SCUBA-2 Positional Offsets}

Surveys such as AS2UDS and ALESS \citep{hodge2013alma} have been motivated by the need to  identify large samples of sub-millimetre galaxies to bright single-dish sub-millimetre sources. In the absence of sub-millimetre interferometry, this  requires statistical associations between the single-dish sources and sources identified in higher-resolution multiwavelength observations.   These
statistical associations are complicated by the significant uncertainties in the single-dish source positions. It is expected that these positional uncertainties are dependant on the SNR of the single-dish detection and in \cite{ivison2007scuba} the ideal theoretical expression for this dependence was given as:
\begin{equation}
   \Delta \alpha = \Delta \delta = 0.6[({\rm SNR})^{2}-(2\beta+4)]^{-1/2}{\rm FWHM},
	\label{eq:ivison}
\end{equation}
where $ \Delta\alpha$ and $\Delta\delta$ are the rms positional errors in R.A.\ and Dec., SNR is the signal to noise ratio of the SCUBA-2 detection, FWHM is the full-width-half-maximum of the single-dish beam, and $\beta$ is the slope of the power law number counts which is required to correct for the Malmquist bias ($\beta\sim-2$).

With our ALMA survey we can empirically test this relation by checking the positional offsets of the AS2UDS sources relative to their corresponding SCUBA-2 source position. We check for a systematic astrometric offset between the parent SCUBA-2 S2CLS position and the detected ALMA sources (the offset corrections from the increased SCUBA-2 integration time for the Cycle 1 sources, mentioned in \S \ref{sec:data}, were pre-applied) by calculating median offsets in Right Ascension and Declination. We find a significant median offset of $-$1$\farcs$6\,$\pm$\,0$\farcs$1 in R.A.\ and $-$0$\farcs$6\,$\pm$\,0$\farcs$1 in Dec., a result of a rounding error in the assigning of phase centres to ALMA and we subsequently apply these corrections to the ALMA phase centres when calculating the radial separations between SCUBA-2 positions and ALMA source detections. We note that this error in phase centre pointings could result in a modest reduction in the number of ALMA counterparts detected at the very edge of the ALMA primary beam, as a shift of this magnitude results in a region with an area corresponding to 12\,\% of the primary beam $\sim$\,8$\arcsec$ from the intended phase centre  falling outside the actual primary beam. We estimate a constraint of $\sim$\,11 potentially missed galaxies due to this offset, calculated by assuming a rotationally symmetric distribution of galaxies around the phase centres and counting the number of galaxies in our catalogue detected at the very edges of our primary beams.

\begin{figure*}
\includegraphics[width=2\columnwidth]{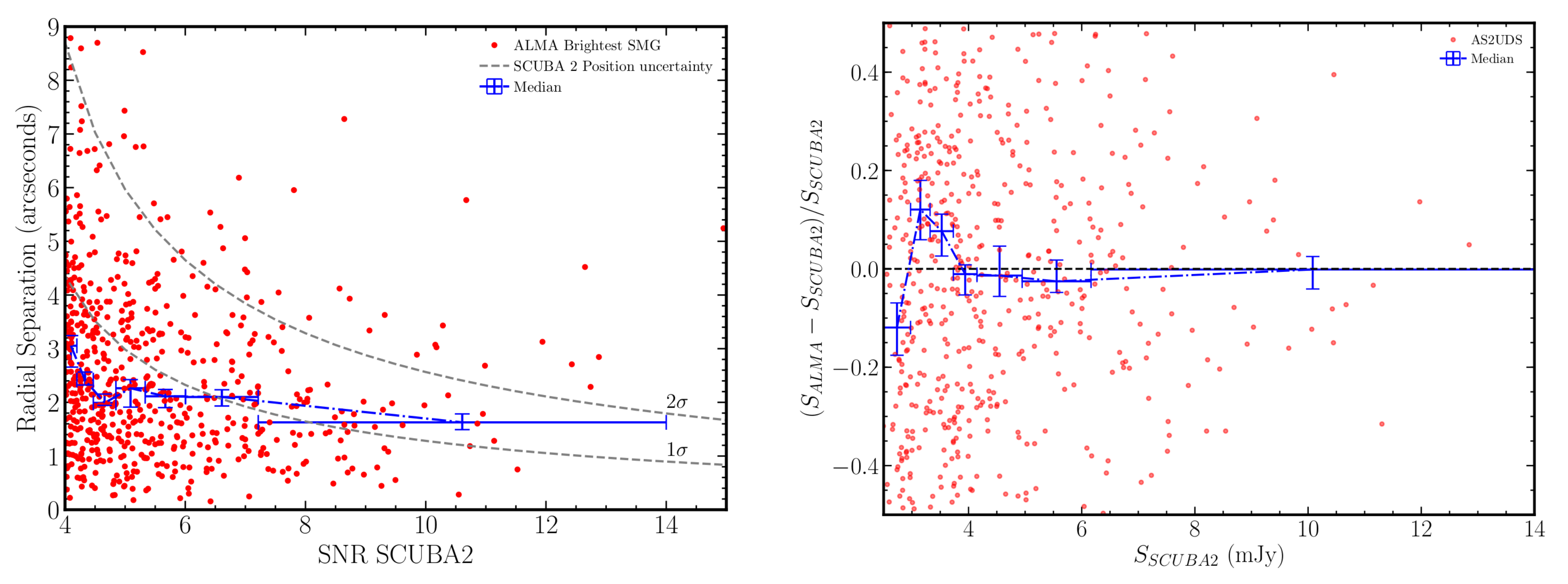}
\caption{\textit{Left:} The offsets between the field centres (corrected for the astrometric offsets and the brightest sub-millimetre detected galaxies in the AS2UDS catalogue. As shown by \citet{simpson2015scuba2} based on our ALMA Cycle 1 pilot programme, these separations are consistent with the predicted uncertainties in source positions for SCUBA-2 detections from \citet{ivison2007scuba} overlaid as the dashed lines. We also plot the median offsets in bins of equal number of brightest sources which show a mostly flat distribution suggesting a fixed search radius for counterparts as opposed to a SNR dependant one would likely be equally effective, most likely due to the influence of secondary components in the maps. \textit{Right:} The recovered flux fraction as a function of original SCUBA-2 flux. Each point is the integrated flux for all sources detected in a field within the primary beam centred on the original S2CLS position. Points above zero indicate maps where we recover a greater flux than the original S2CLS detection, a result of either noise boosting or sources being detected near the edge of the map thus are primary beam corrected to higher fluxes, and points below zero are maps where we failed to recover all the flux from the parent single-dish source. Overlaid is the median in bins of equal number of ALMA maps that show that for SCUBA-2 sources above $S_{850}\sim3.5$\,mJy we on average recover all the single-dish flux in our ALMA maps.}
\label{fig:offsets}
\end{figure*}

In Figure~\ref{fig:offsets} we show the radial offsets as a function of the SNR of the parent single-dish source and overlay the predicted uncertainties on the single-dish positions from \cite{ivison2007scuba}. In agreement with the predicted uncertainties, 63\,$\pm$\,3\,\% of the brightest detected sources are within the 1-$\sigma$ uncertainty and 92\,$\pm$\,4\,\% within the 2-$\sigma$ uncertainty. However, the actual functional form of the uncertainty in position with signal-to-noise ratio has previously been used to determine the search radius for counterparts \citep[e.g.][]{biggs2011laboca} and therefore we look in more detail at how the trend of  median offsets varies as a function of signal-to-noise in Figure~\ref{fig:offsets}. The median offsets shows a much flatter trend in comparison to the theoretical expectation, with the faintest SCUBA-2 sources (SNR\,$<$\,5) median offsets lying significantly below the 1-$\sigma$ theoretical prediction of \cite{ivison2007scuba}, whilst the median for the brighter (SNR\,$>$\,6) sources lies above. This trend suggests that altering the search radius to identify counterparts based on signal-to-noise could result in a failure to correctly identify the correct counterparts in high significance SCUBA-2 detections. This failure to follow the expected behaviourt is likely due to the presence of more than one sub-millimetre galaxy in a map of a single-dish source, which is not accounted for by the model, such `multiplicity' is particularly prevalent in brighter sources \citep[e.g.][]{stach2018alma}.

%
%

\subsection{Flux Recovery}

To determine the reliability of our flux measurements we next compare the total flux recovered from all sources detected in an ALMA map against the corresponding SCUBA-2 flux. To do this we correct the SCUBA-2 850\,$\mu$m fluxes to the ALMA wavelength using a $S_{870}/S_{850}=0.95$ factor derived from the ALESS survey composite spectral energy distribution (SED) \citep{swinbank2014alma} redshifted to $z=2.5$ (Figure.~\ref{fig:offsets}). For SCUBA-2 sources brighter than $S_{850}\geq$\,4\,mJy\,beam$^{-1}$ we have a median recovery rate of 97$^{+1}_{-2}$\,\% of the flux in the ALMA pointings, which is well within the systematic flux uncertainties for both SCUBA-2 \citep[5--10\,\%,][]{dempsey2013scuba} and ALMA. Therefore we can be confident that all significant contributors to the sub-millimetre fluxes are successfully recovered in this flux regime. However the binned median flux recovery is 87$^{+5}_{-8}$\,\% for $S_{850}=2.9$--$3.1$\,mJy\,beam$^{-1}$ and 71$\pm$6\,\% at $S_{850}=2.5$--$2.9$\,mJy\,beam$^{-1}$. These medians are biased low by the blank maps with no detected ALMA sources. If we exclude those maps, this results in an 88$\pm$6\,\% flux recovery at 2.5$\leq S_{850}\leq$\,2.9\,mJy\,beam$^{-1}$ and 112\,\% flux recovery at 2.9$\leq S_{\rm 850}\leq$\,3.1\,mJy\,beam$^{-1}$. Therefore only at the faintest SCUBA-2 fluxes do we see the suggestion of decline in the recovered fraction of the SCUBA-2 flux, and in Section \ref{sec:blanks} we show that for the majority of cases this is likely due to faint sub-millimetre galaxies just below our 4.3\,$\sigma$ threshold being present in the ALMA maps.

\subsection{Photometric Redshifts}

With the final AS2UDS catalogue matched to the extensive multiwavelength coverage in the UDS field we derive the multiwavelength properties for our SMGs from SED fitting from a maximum of 22 filters ($U, B, V, R, I, z, Y, J, H, K,$ \emph{IRAC} 3.6, 4.5, 5.8, and 8.0\,$\mu$m, MIPS 24\,$\mu$m, PACS 100, and 160\,$\mu$m, SPIRE 250, 350, and 500\,$\mu$m, $S_{870}$, and $S_{\rm 1.4\,GHz}$) using \textsc{magphys} \citep{da2008simple}. \textsc{magphys} employs the stellar population synthesis models of \citet{bruzual2003stellar} with a \citet{chabrier2003galactic} IMF, combined with a two-component description of the dust attenuation in the ISM and stellar birth clouds \citet{charlot2000simple} in an energy-balance model to ensure consistency between the mid- to far-infrared emission from dust re-processing of the stellar emission and the integrated (dust-attenuated) stellar emission of the galaxy \citep[details in:][]{da2008simple}. \citet{da2015alma} extended this method to include the computation of photometric redshifts simultaneously with the constraints on other physical parameters when fitting the SEDs of dusty star-forming galaxies. We use that version of the code in this paper (for a full description of \textsc{magphys}-photoz and code release see also Battisti et al. in prep). To test the reliability of the predicted photometric redshifts, \textsc{magphys} was fitted to 14 photometric bands of 7316 spectroscopic sources from the UKIDSS UDS DR11 photometric catalogue. The median relative difference, $\Delta z = (z_{\rm spec}-z_{\textsc{magphys}})/(1+z_{\rm spec}$), was found to be -0.0056 with a standard deviation of 0.30. Similarly, for the 44 AS2UDS SMGs with spectroscopic redshifts we find a median $\Delta z=-0.02$ with a standard deviation of 0.25. The complete description of our \textsc{magphys} SED fitting and the resulting multiwavelength properties is described in U. Dudzevi\v{c}i\={u}t\.{e} et al (in prep.).

\section{Results and Discussion} \label{sec:results}

In this section we first catalogue the detected sources in the ALMA maps and discuss the trends with redshifts in our sample and the potential causes for maps which lack detected sources. We also study the properties of those in the sample hosting actively accreting super-massive black holes (SMBHs) and the connection of these galaxies to the formation of massive, passive galaxies at high redshifts.


\subsection{AS2UDS Catalogue}
The complete AS2UDS catalogue identifies 708 sub-millimetre galaxies brighter than $S_{870}=$\,0.58\,mJy from the original 716 SCUBA-2 sub-millimetre sources, roughly five times larger than the \cite{miettinen2017alma} study in COSMOS or ALESS \citep{hodge2013alma}. In Table \ref{tab:catalogue} we present the AS2UDS catalogue.

From the 716 ALMA maps: one contains four SMG detections, six have three SMG detections, 78 include two, and 530 detect just a single SMG above 4.3\,$\sigma$. The majority of the maps containing multiple SMGs correspond to the brighter SCUBA-2 sources, so as  \citet{stach2018alma} showed the rate of occurrence of  multiple counterparts is 26\,$\pm$\,2\,\% in SCUBA-2 sources with fluxes brighter than $S_{850}\geq$\,5\,mJy and 44\,$\pm$\,16\,\% at $S_{850}\geq$\,9\,mJy.  The presence of these multiple strongly star-forming galaxies in close proximity  may be hinting at a role for major mergers in driving  the enhanced star-formation rates in some of these systems. Indeed, the small subset of AS2UDS covered by the high-resolution {\it HST} imaging (see Figure \ref{fig:candels}) indicates that  many of the SMGs are morphologically complex, with close companions and/or structured dust obscuration, consistent with the \emph{HST} imaging for the ALESS survey \citep{chen2015alma}. From visual inspection, independently carried out by two of the authors, the CANDELS coverage suggests 50$\pm$10\,\% of our SMGs are either clear mergers or are disks with likely companions with similar colours on $<$20\,kpc scales, and we group these two visual classifications as `likely interacting'. The median redshifts for the `likely interacting' SMGs is $z_{\rm phot}=2.2\pm0.1$, significantly lower than the median redshift for the whole SMG sample (see \S \ref{sec:photzcat}). Indeed, if we cut our sample to SMGs at redshifts $z_{\rm phot}<2.75$, the redshift range at which we could reasonably expect to detect interactions in the CANDELS imaging then this `likely interacting' classification increases to 17/21 or $\sim$80\,\% of our SMGs, again this is consistent with \citet{chen2015alma}. This small subset in the CANDELS imaging strongly hints that major mergers are playing a role in driving the enhanced star-formation rates that result in our galaxies being selected in the sub-millimetre maps.

\begin{table*}
 \caption{The AS2UDS Catalogue.  The ALMA\_ID comprises a 4-digit numbers giving the field identifier from the parent S2CLS survey and a final digit which  identifies the individual galaxies detected from each S2CLS source with ascending numbers for descending flux. The source co-ordinates are the source centroids from \textsc{sextractor} from the $uv$-tapered detection maps. Field rms is the sigma-clipped rms across the entire map and are plotted in Figure \ref{fig:rms}. PB Correction gives the primary beam-correction factor at each source location, a higher number indicating galaxies further away from the ALMA phase centre. The S2CLS Flux columns lists the flux-deboosted 850\,$\mu$m flux from \citet{geach2017scuba}. The Aper SNR column is the 0$\farcs$5 diameter aperture flux SNR used to select of galaxies in the catalogue and the Final Flux is the 1$\farcs$0 diameter aperture flux of the galaxy with the primary beam, aperture, and flux boosting corrections applied. The complete catalogue also contains information on which ALMA cycles the galaxies were observed, the individual flux corrections and the restoring beam properties of the maps.}
 \label{tab:catalogue}
 \begin{tabular}{cccccccc}
  \hline
  ALMA\_ID & Source\_RA & Source\_Dec & Field RMS & PB Correction & S2CLS Flux & Aper SNR & ALMA Final Flux\\
 & J2000 & J2000 & (mJy\,beam$^{-1}$) & & (mJy) & & (mJy) \\
  \hline
    AS2UDS0001.0 & 34.62777 & $-$5.52545 & 0.38 & 1.01 & 52.6$\pm$0.0 & 35.7 & 30.2$\pm$0.86\\ 
    ... & ... & ... & ... & ... & ... & ... & ...\\
  \hline
 \end{tabular}
\end{table*}

\begin{figure*}
\includegraphics[width=2\columnwidth]{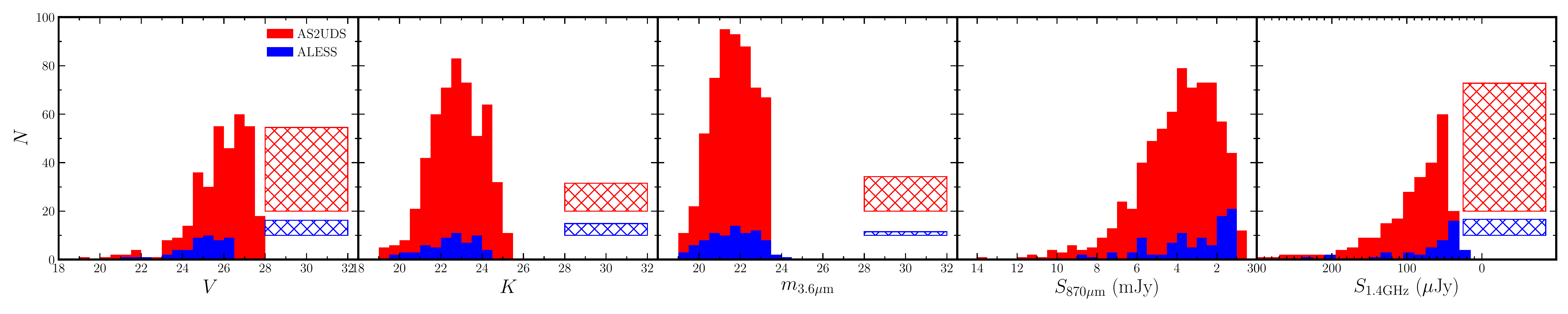}
\caption{Apparent magnitude distributions in the $V$, $K$ and 3.6\,$\mu$m wavebands and the flux densities in 870\,$\mu$m and 1.4\,GHz for AS2UDS.  For comparison we show the same distributions for the earlier ALESS survey, and in both cases the hatched regions showing the non-detections. The median magnitudes/fluxes for  AS2UDS  are $V=$\,26.1\,$\pm$\,0.1, $K=$\,22.8\,$\pm$\,0.1, $m_{\rm 3.6}=$\,21.65$^{+0.06}_{-0.03}$, $S_{870}=$\,3.73$^{+0.03}_{-0.10}$\,mJy, and $S_{\rm 1.4\,GHz}=$\,86$^{+2}_{-5}$\,$\mu$Jy, with the estimates for the smaller ALESS survey in good agreement. These illustrate that typical SMGs have  red optical-near infrared colours and so  are very faint in the optical with a large fraction of the population undetected in the bluest wavebands.  In contrast, most of the SMGs are detected
at 3.6\,$\mu$m. }
\label{fig:magdist}
\end{figure*}

In Figure \ref{fig:magdist} we show the apparent magnitude/flux density distributions for the AS2UDS SMGs in  $V$, $K$, $m_{\rm 3.6}$, $S_{870}$, and 1.4\,GHz bands and for comparison the corresponding distributions of the ALESS SMGs \citep{hodge2013alma,simpson2014alma}. The median ALMA flux density for the AS2UDS SMGs is $S_{870}=$\,3.73$^{+0.03}_{-0.10}$\,mJy, very similar to the $S_{870}=$\,3.5\,$\pm$\,0.3\,mJy for ALESS. Of the 708 SMGs, 529 (75\%) have $K$-band counterparts in the UKIDSS-UDS DR11 catalogue to $K\leq$\,25.9\,mag.  However,  excluding ALMA SMGs which fall either outside the DR11 WFCAM $K$-band image or in regions which are flagged as  being shallower, the detection rate corresponds to 84\,$\pm$\,4\,\% (484/577). The median apparent magnitudes for these AS2UDS SMGs are $V=$\,26.1\,$\pm$\,0.1, $K=$\,22.8\,$\pm$\,0.1, $m_{\rm 3.6}=$\,21.65$^{+0.06}_{-0.03}$. These values are all in good agreement with the equivalent measurements for the ALESS sample of $V=$\,26.1$^{+0.2}_{-0.1}$,  $K=$\,23.0$^{+0.3}_{-0.1}$, and $m_{\rm 3.6}=$\,21.8$\pm$0.2. Hence we conclude that these two ALMA follow-up surveys of single-dish detections find SMGs with multi-wavelength magnitude distributions in excellent agreement with each other. These photometric properties demonstrate the dusty nature of the target populations with an increasing number of non-detections towards the bluer optical wavebands, e.g. only \ 64\,$\pm$\,2\,\% of $K$-band detected SMGs are detected in the $V$-band brighter than $V<$\,27.8. We note that, unlike in the pilot AS2UDS sample \citep{simpson2015scuba2}, in the full sample there is no evidence that a lack of a $K$-band detection is a function of the $S_{870}$, with a median ALMA flux of $S_{870}=$\,3.65$^{+0.12}_{-0.08}$\,mJy for the $K$-detected SMGs and $S_{870}=$\,3.5$^{+0.2}_{-0.1}$\,mJy for the $K$-undetected SMGs, the same trend was found in \citet{cowie2018submillimeter}.

\subsection{Redshift Distribution and Trends} \label{sec:photzcat}

\begin{figure}
\includegraphics[width=\columnwidth]{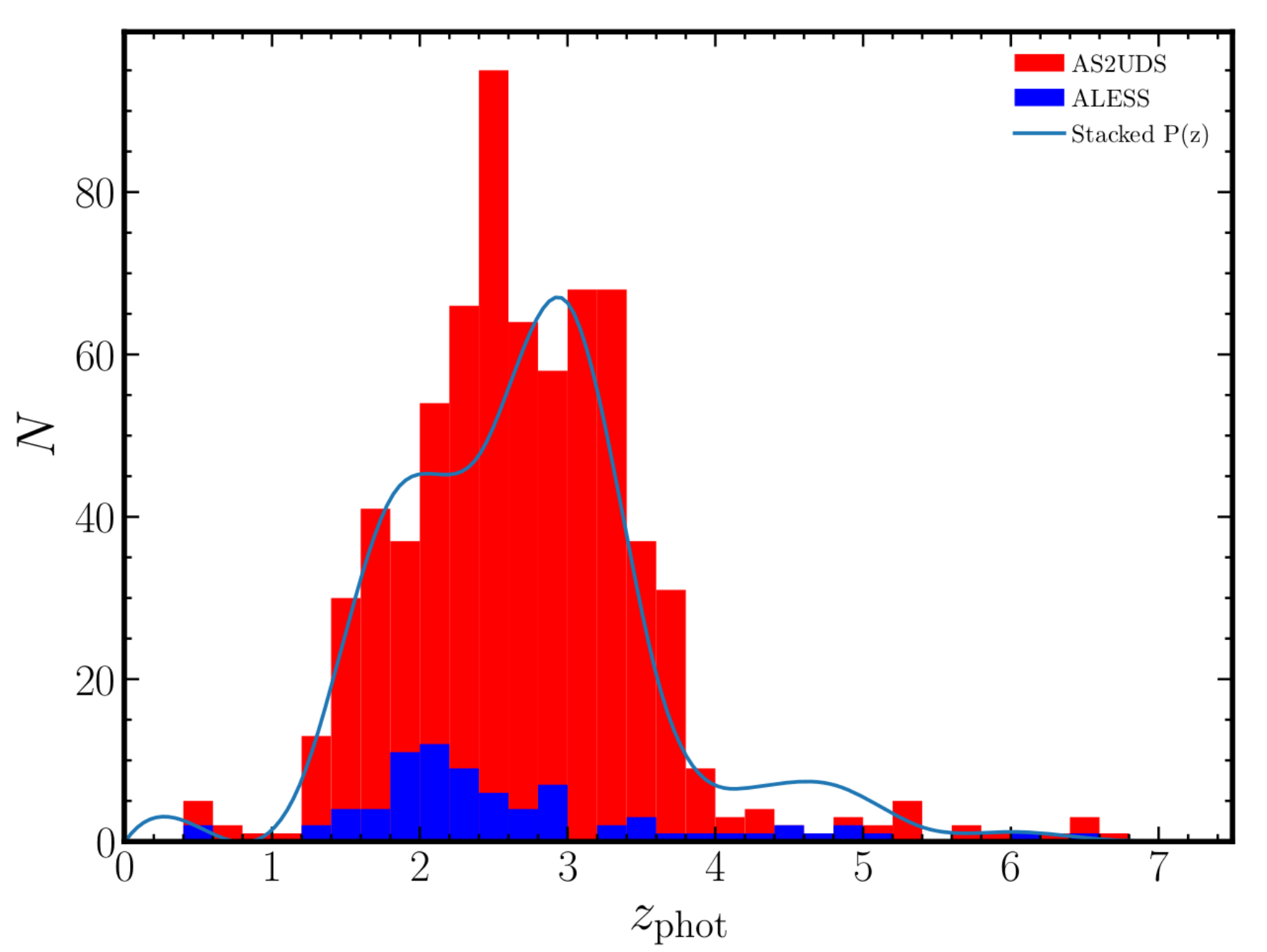}
\caption{The photometric redshift distribution for the AS2UDS SMGs based on \textsc{magphys} analysis (Dudzevi\v{c}i\={u}t\.{e} et al.\ in prep). We determine a median redshift of $z_{\rm phot}=$\,2.61\,$\pm$\,0.09, consistent with the ALESS photometric redshift distribution \citep{da2015alma}, which has a median $z_{\rm phot}=$\,2.7\,$\pm$\,0.1. The ALESS distributions also display a  fraction of sources at $z_{\rm phot}>$\,3 galaxies,  38$^{+7}_{-6}$\%, that agrees with that derived here for our large AS2UDS sample,  33$^{+3}_{-2}$\,\%. To illustrate the influence of the asymmetric redshift uncertainties we also overlay the stacked probability distributions for the photometric redshifts of all AS2UDS SMGs from \textsc{magphys}.  This closely resembles the median photometric redshift distribution, suggesting that the full combined  photometric redshift is not sensitive to any secondary redshift solutions in individual sources.}
\label{fig:rdist}
\end{figure}

The photometric redshift distribution for all 708 SMGs is shown in Figure~\ref{fig:rdist}.  We derive a median redshift of $z_{\rm phot}=$\,2.61\,$\pm$\,0.09 and a tail at higher redshifts, with 33$^{+3}_{-2}$\,\% of galaxies at $z_{\rm phot}>$\,3. This median redshift is consistent with the  Cycle 1 pilot sample of bright SMGs from \citet{simpson2015scuba2} of $z_{\rm phot}=$\,2.65\,$\pm$\,0.13 (or $z_{\rm phot}=$\,2.9\,$\pm$\,0.2 including optically faint SMGs without precise photometric redshifts). The more direct comparison to our sample are the \textsc{magphys}-derived redshift distribution for ALESS from \cite{da2015alma} who found a median redshift of $z=$\,2.7\,$\pm$\,0.1 with a similar fraction of galaxies in the $z_{\rm phot}>$\,3 tail of 38$^{+7}_{-6}$\%.  

%
%

We next compare our photometric redshift distribution with previous spectroscopic redshift surveys selected at 850\,$\mu$m SMGs. \cite{chapman2005redshift} found a median redshift for their 73 radio-detected SMGs of $z=$\,2.2\,$\pm$\,0.1 ($z=$\,2.3\,$\pm$\,0.2 after statistically correcting for the radio bias). The lower median redshifts  is unsurprising given the lack of a negative $K$-correction at radio wavelengths.  We can confirm this by looking at the AS2UDS SMGs with radio bright counterparts where we find a median redshift of $z=$\,2.4$^{+0.3}_{-0.9}$, which is in reasonable  agreement with the \cite{chapman2005redshift} sample. The spectroscopic sample of \cite{danielson2017alma}, which consisted of 52 ALMA-confirmed SMGs detected in the ALESS catalogue found a median redshift of $z=$\,2.40\,$\pm$\,0.10, in rough agreement with the photometric redshift distribution of AS2UDS, however, this spectroscopic redshift sample is biased to optically and near-infrared bright counterparts.

There have also been wide-field single-dish surveys at longer wavelengths, $\sim$\,1--1.2\,mm, which are now being similarly followed up with sub/millimetre interferometers \citep[e.g.][]{smolvcic2012millimeter,miettinen2017alma}. In particular, \citet{brisbin2017alma} obtained ALMA interferometric observations of 129 1.25\,mm sources in COSMOS and determined a median redshift (from a heterogeneous mix of spectroscopic, photometric and colour-based estimates or limits) of $z\sim$\,2.48\,$\pm$\,0.05 for a sample with an equivalent 870-$\mu$m flux of $S_{870}\gtrsim$\,6\,mJy. This is in reasonable agreement with our median redshift, given the different flux and waveband selections.

We can also compare our median redshift with those derived for the typically fainter samples of SMGs from `blank field' ALMA surveys \citep[e.g.][]{walter2016alma,dunlop2016deep}. The ALMA `blank field' surveys are usually undertaken at longer wavelengths than our 870-$\mu$m observations, normally in Band 6 at $\sim$\,1.1--1.3\,mm, to exploit the larger ALMA primary beam to increase the area coverage.  Three ALMA surveys of blank fields have reported median redshifts for their samples, although we caution that these all study fields in the GOODS-South area and so are not truly independent. \citet{aravena2016alma} report a median redshift of $z=$\,1.7\,$\pm$\,0.4 for a sample of nine galaxies with an 870-$\mu$m equivalent flux limit of $S_{870}\gs$\,0.1\,mJy, \citet{dunlop2016deep} report $z=$\,2.0\,$\pm$\,0.3 for 16 sources with optical/near-infrared counterparts and $S_{\rm 870}\gs$\,0.3\,mJy and \citet{franco2018goods} find $z=$\,2.9\,$\pm$\,0.2 for a sample of 20 galaxies with $S_{870}\gs$\,2\,mJy. These studies are challenging and the results from all of them are currently limited by  statistics, but there is a hint that the deeper surveys are finding lower median redshifts than from our bright sample, in the same sense as the trend we discuss below.

We next study the variation in median redshift with 870-$\mu$m flux density within AS2UDS. It has long been claimed that there is variation in redshift with  sub-millimetre flux, in the sense that the more luminous sub-millimetre sources typically lie at higher redshifts. This behaviour was first seen in powerful high-redshift radio galaxies \citet{archibald2001submillimetre}, but the existence of a similar trend has been claimed for sub-millimetre galaxies, although these claims have typically suffered from incomplete, heterogeneous or highly uncertain redshift estimates \citep[e.g.][]{ivison2002deep,pope2005hubble,younger2007evidence,smolvcic2012millimeter,koprowski2014reassessment,brisbin2017alma}. The initial search for a $z$--$S_{870}$ trend in the ALESS sample \citep{simpson2014alma} and in the initial bright pilot for AS2UDS \citep{simpson2015scuba2} both found weak  trends.  Moreover, these were based on relatively small and incomplete samples owing to the reliance on optical-near-infrared photometric redshifts.  When allowance was made for this incompleteness, the trends were not statistically significant.

One benefit of using {\sc magphys} for our photometric redshift analysis is that we can obtain more complete and homogeneous redshifts estimates, not only for those  sources detected in the optical to near-infrared bands, but also where only longer wavelength constraints are available.  So, we begin by determining if the SMGs which are undetected in the $K$-band are at higher redshift than those which have a $K$-band counterpart, as would be required for their to be a trend of $z$--$S_{870}$.  For galaxies with $K$-band counterparts we find a median $z_{\rm phot} = $\,2.6\,$\pm$\,0.1 and for the $K$-band non-detections we find a median $z_{\rm phot} = $\,3.0\,$\pm$\,0.1, suggesting that indeed the population of $K$-faint SMGs, whilst not deviating dramatically in distribution of $S_{\rm 870}$ as mentioned above, do represent a higher redshift subset of SMGs.

\begin{figure}
\includegraphics[width=\columnwidth]{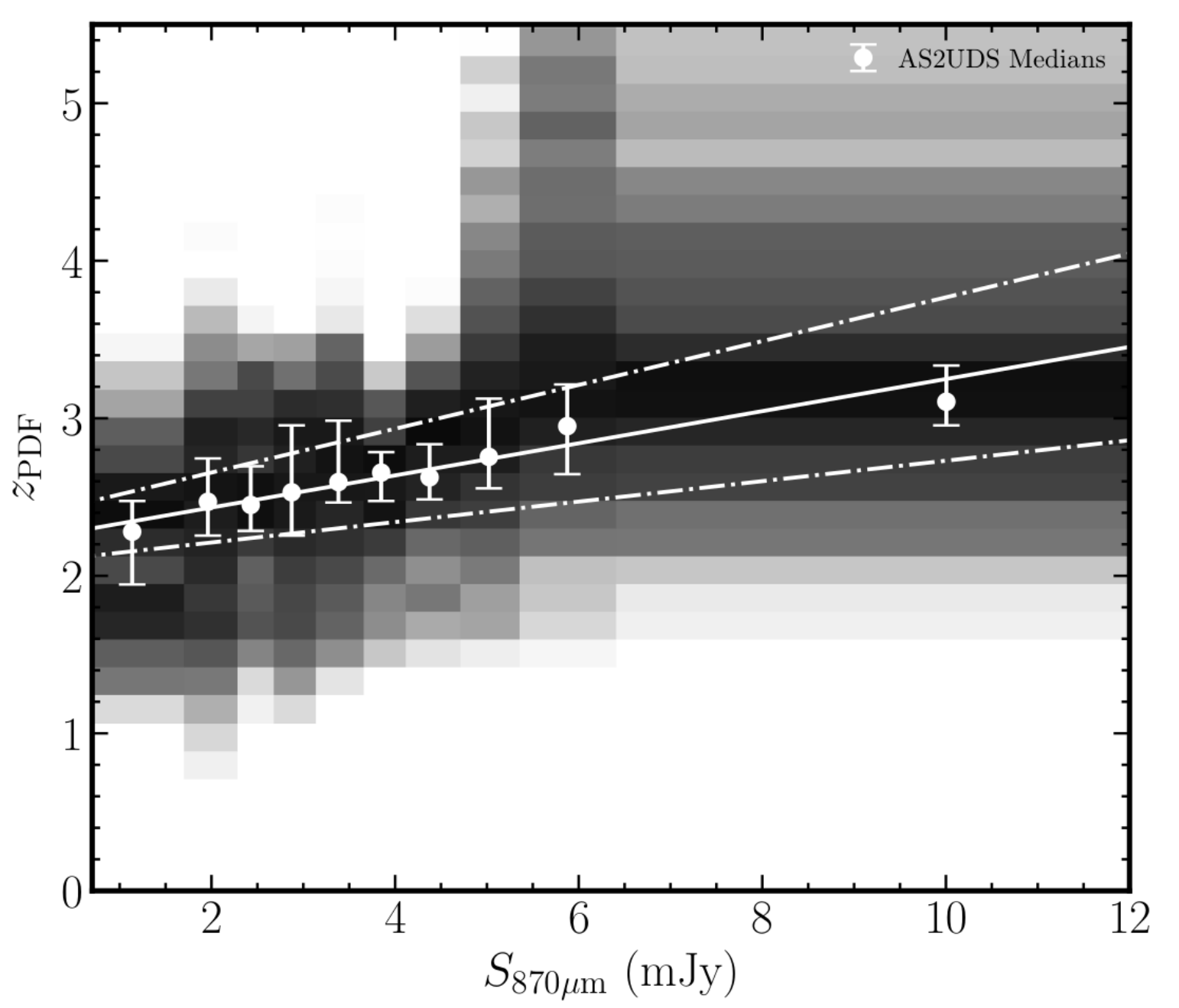}
\caption{The trend between photometric redshifts and ALMA 870\,$\mu$m flux density for AS2UDS SMGs. We bin the galaxies by ALMA 870\,$\mu$m flux density with bins of equal galaxy numbers and find that the median redshift and flux of each bin.  These show a significant trend of increasing redshift with increase flux. The linear fit to this trend has a gradient of 0.09\,$\pm$\,0.01\,mJy$^{-1}$ and we plot this as the solid line and the 3-$\sigma$ errors for this fit as the dashed lines. In addition to the medians of the bins, the greyscale background shows the stacks of the redshift probability density functions for each bin, which also display the same flux density evolution.}
\label{fig:fluxvz}
\end{figure}

In Figure \ref{fig:fluxvz} we show the photometric redshifts as a function of the ALMA 870\,$\mu$m flux density for AS2UDS using our   complete {\sc magphys}-derived photometric redshifts.  We over plot a linear fit to the median redshifts in bins of equal number of galaxies. This fit shows a  trend of median redshift with 870\,$\mu$m flux density with a highly significant gradient of 0.09\,$\pm$\,0.02\,mJy$^{-1}$.   The significance of this result is perhaps surprising given the absence of a strong trend in either ALESS \citep{simpson2014alma} or our Cycle 1 pilot programme \citep{simpson2015scuba2}. However, the most likely explanation is simply the much smaller sample sizes ($n<$\,100) in those studies. To test this  we randomly draw samples of 100 galaxies from the complete AS2UDS catalogue 500 times and repeat the same fitting procedure.  In this case we recover an average gradient of 0.08\,$\pm$\,0.06\,mJy$^{-1}$ confirming that the smaller samples employed in all previous tests mean that it would be impossible to reliably recover the trend we see at higher than  $\sim$\,1.5\,$\sigma$ significance. In addition, to confirm that our detection incompleteness isn't driving this trend we perform the same trend fitting to the galaxies with fluxes S$_{870}>$\,4\,mJy, again finding a similar gradient of 0.08\,$\pm$\,0.03\,mJy$^{-1}$.
  
%
%

What is the physical process responsible for the $z$--$S_{870}$ trend we see in AS2UDS? Due to the negative $K$-correction of the dust SEDs in typical SMGs, across the redshift range ($z\sim$\,1.5--5) that we measure this trend, the population have a roughly constant observed flux density at a fixed luminosity (and temperature). Hence the brighter galaxies found at the higher redshifts are intrinsically more luminous. The  trend we see then suggests there is a strong luminosity evolution for our SMGs out past $z >$\,3, with the most far-infrared luminous galaxies being found in greater numbers in the early Universe. This could be symptomatic of galaxy `downsizing' \citep{cowie1996new}, i.e.\ the more massive galaxies are forming at higher redshifts. We will return to the issue of potential evidence of 'downsizing' on a forthcoming paper on the clustering of the AS2UDS SMGs in the UDS field (Stach et al.\ in prep.).

Finally, to attempt to model the variation in median redshifts, we employ the \citet{bethermin2015influence} models of SMG number counts and redshift distributions to model the median redshifts of surveys from two variables; the wavelength of selection of the SMGs and the flux density depth of the survey. The \citet{bethermin2015influence} model suggests that surveys at longer wavelengths will recover a higher median redshift and that, due to the luminosity evolution, the fainter sources found in deeper surveys will predominantly lie at lower redshifts. Whilst this phenomenological model reproduces the broad trend we see in AS2UDS, as well as the comparatively low median redshifts of \citet{aravena2016alma} and \citet{dunlop2016deep} due to their fainter flux limits, it does under-predict the median redshift for our survey. For a $S_{870}>$\,4\,mJy flux limited sample the model predicts $z\sim$\,2.6 compared to our $z = $\,2.8\,$\pm$\,0.1, and this $\Delta z\sim-$0.2 offset between the model and our survey roughly persists across any choice of faint flux limit in our catalogue, indicating that further tuning of this phenomonological model may be needed.

\subsection{Blank Maps} \label{sec:blanks}

\begin{figure*}
\includegraphics[width=2\columnwidth]{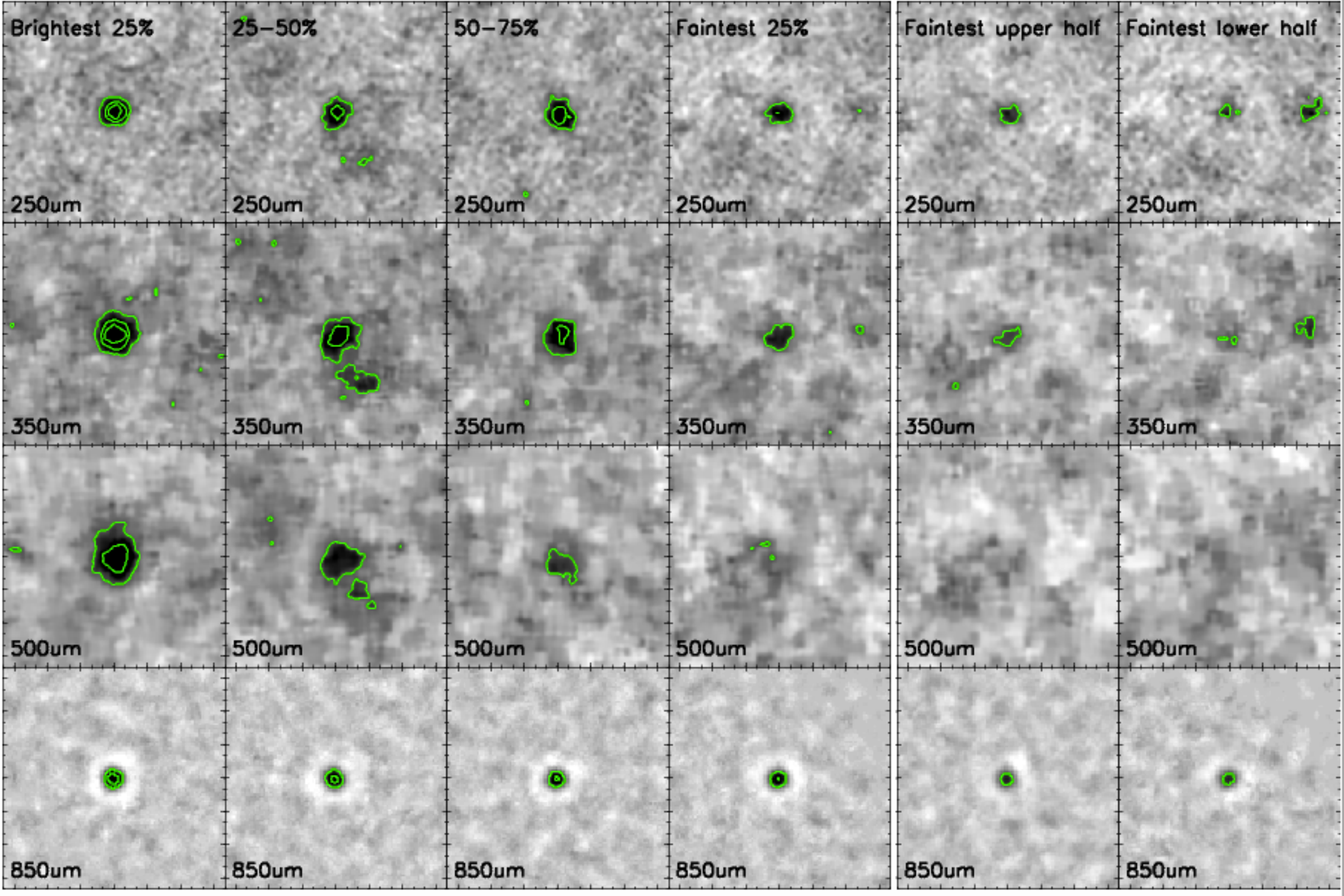}
\caption{\textit{Herschel}/SPIRE and SCUBA-2 stacks on the 101 ALMA `blank' maps in four bins of original SCUBA-2 flux. Even in the faintest SCUBA-2 bin we detect emission at 250\,$\mu$m and 350\,$\mu$m. Further splitting the faintest bin into two results in detected emission strongly suggesting these `blank' maps are not typically the result of spurious single-dish SCUBA-2 detections. Each map is 120$\arcsec$ square and is centred on the SCUBA-2 source position.}
\label{fig:stacks}
\end{figure*}

There are 101 ALMA maps of SCUBA-2 sources in which we find  no ALMA detected source above our 4.3\,$\sigma$ detection threshold, these are
termed `blank' maps. This rate of non-detection is considerably higher than the expected 2\,\% false-positive rate predicted from \cite{geach2017scuba} for the initial SCUBA-2 catalogue. Whilst these `blank' maps  typically correspond to the fainter single-dish sources: the median SCUBA-2 flux of the `blank' maps is $S_{\rm 850}=$\,4.0\,$\pm$\,0.1\,mJy, compared to $S_{850}=$\,4.5\,$\pm$\,0.1\,mJy for the whole sample, the SCUBA-2 noise properties of the `blank' maps are consistent with those of the maps with detections. To confirm that these are not dominated by false positive detections in the parent SCUBA-2 catalogue we stack the \textit{Herschel}/SPIRE and SCUBA-2 maps at the locations of the 101 `blank' map sources ranked in four bins of their SCUBA-2 flux and further sub-dividing the faintest quartile in SCUBA-2 flux into two independant halves, shown in Figure.~\ref{fig:stacks}.

%
%

\begin{figure*}
\includegraphics[width=2\columnwidth]{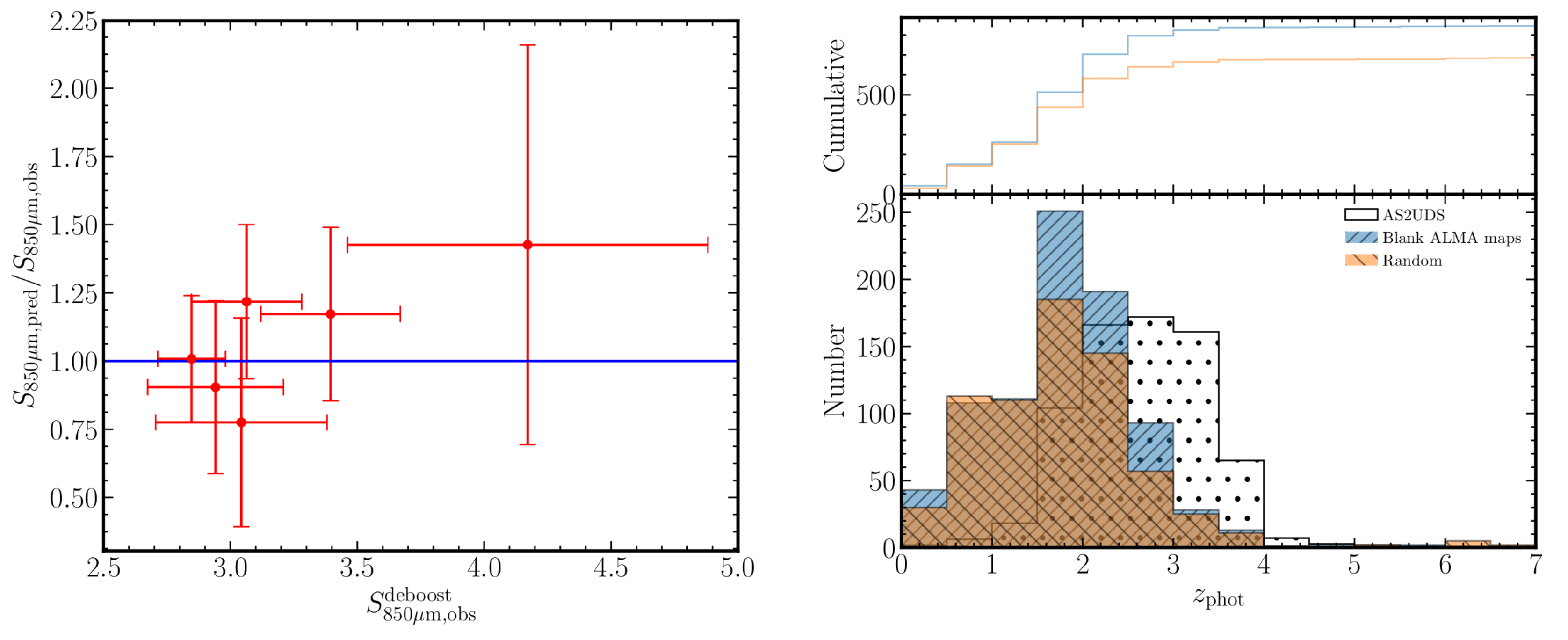}
\caption{\textit{Left:} Predicted 850\,$\mu$m flux from fitting SEDs to the stacked \textit{Herschel}/SPIRE fluxes of sub-samples of SCUBA-2 sources,  ranked in terms  of SCUBA-2 850\,$\mu$m flux,  where our ALMA observations detected no SMGs.  These are compared to the observed SCUBA-2 850\,$\mu$m flux for the equivalent sub-sample.  On average we recover the predicted SCUBA-2 fluxes from the SED fitting which strongly suggests that the majority of these SCUBA-2 sources are real, rather than spurious detections in the parent catalogue. \textit{Right:} The \textsc{magphys} photometric redshift distribution of the $K$-band DR11 UKIDSS UDS sources within the primary beam  of the ALMA `blank' maps.
This is compared to an identical sized area  randomly distributed across the UDS field ('random'). We see a significant excess of galaxies
in the `blank' map regions compared to a random field, corresponding to an excess of 153 galaxies at $z_{\rm phot}=$\,1.5--4. This suggests
that these apparently `blank' maps might each contain 1--2 faint SMGs lying at redshifts similar to the distribution of the brighter, detected, SMGs which is also shown.}
\label{fig:blanks}
\end{figure*}

For each subsample we detect emission in the \textit{Herschel}/SPIRE at 250 and 350\,$\mu$m and even when further sub-diving the faintest subsample into two there is still emission detected at 250\,$\mu$m strongly suggesting these ALMA-blank maps are not just the result of spurious single-dish sources.  To compare  the 850\,$\mu$m fluxes implied by these SPIRE detections with the original SCUBA-2 measurements, we predict 850\,$\mu$m flux from the SPIRE stacks by fitting a modified blackbody SED with $z=$\,2.5, a dust temperature $T_{d}=$\,35\,K, and emissivity $\beta=$\,1.5 to each of the 250/350/500\,$\mu$m stacked fluxes. In Figure~\ref{fig:blanks} we show the predicted 850\,$\mu$m flux from these SED fits against the original detected 850\,$\mu$m emission from SCUBA-2. Whilst there is significant scatter in these estimates, the 850-$\mu$m fluxes predicted from the extrapolation of the stacked \textit{Herschel} fluxes is consistent with that observed by SCUBA-2.   We view this as a strong confirmation that the majority of these SCUBA-2 sources with ALMA `blank' maps are not spurious detections in the original S2CLS catalogue.
  
%
%

We conclude that spurious single-dish detections are not the dominant cause for `blank' maps, therefore, we next examine possibilities for missing ALMA counterparts in our ALMA maps of these sources. An alternative possibility is  multiplicity \citep[e.g.][]{karim2013alma}, where a single-dish source splits into more than one, fainter galaxies. The  combined flux from these galaxies would recover the single-dish flux, but individually each galaxy is below our detection threshold. To check if this is a possibility we look for over-densities of candidate $K$-band detected galaxies in our ALMA blank maps in comparison to a `random' location within the UKIDSS UDS coverage. In Figure \ref{fig:blanks} we show the \textsc{magphys} photometric redshift distributions for both the $K$-band detected sources within the primary beams of all 101 `blank' maps  and for a `random' sample covering the same area as the 101 ALMA primary beams, but randomly distributed across the UDS field. We can see an excess of $K$-selected galaxies in the ALMA-blank maps.  The majority of this excess arises from galaxies at redshifts of $z\sim$\,1.5--4, 
corresponding to the redshift range where our detected SMG population peaks, 
with 153 excess galaxies in this redshift range in the `blank' map regions compared to the `random' area (a factor of 1.36$^{+0.13}_{-0.12}$ increase). This over-density comprises an average excess in a `blank' map of 1.5$\pm$0.5 $K$-band sources (within the expected sub-millimetre galaxy redshift range).   Stacking the ALMA emission of all of the galaxies at $z=$\,1.5--4 in these `blank' maps recovers an average flux
of $S_{870}=$\,0.12\,$\pm$\,0.02\,mJy corresponding to an average flux per map of $S_{870}=$\,0.7\,$\pm$\,0.1\,mJy to be split between
the $\sim$\,1.5 excess sub-millimetre-bright galaxies in these regions, suggesting a typical  flux of $S_{870}\sim$\,0.5\,mJy.
This excess of sources and detected sub-millimetre flux is consistent with the interpretation of the `blank' maps as resulting from the single-dish source comprising flux more than one, faint, sub-millimetre galaxy below our detection threshold.

To test this, in Cycle 5 we re-observed ten of the brightest S2CLS sources which returned no ALMA detections in our Cycle 3 and 4 maps. To both  test for the presence of multiple faint SMGs in these fields and to eliminate the possibility of non-detection due to source flux being resolved out in the interferometric images, these observations were much deeper ($\sigma_{870}=$\,0.085\,mJy\,beam$^{-1}$) and at much  lower resolution than the original maps (synthesised beam: 0$\farcs$81$\times$0$\farcs$54). From the ten single-dish sources we detect 16 $>$\,4.3\,$\sigma$ SMGs.  We find four SMGs in the ALMA map of a single SCUBA-2 source which had previously had a `blank' map, two SMGs in each of a further four maps, four SCUBA-2 sources which have only a single corresponding SMG and just one ALMA `blank' map which remains blank in these deeper observations   (UDS0101, which may be a true false-positive in the S2CLS catalogue).    These deeper observations thus confirm that multiplicity is a significant driver of the `blank' maps and that it remains an issue even for faint SCUBA-2 sources.

The ALMA galaxies in these re-observed `blank' maps, as expected, are amongst the faintest in the catalogue, with a median deboosted flux of 1.13$^{+0.44}_{-0.16}$\,mJy.  Even at the $\sim $\,4$\times$ depth of these observations, and with the high detection rate of SMG counterparts, we still only recover an average of 52$^{+5}_{-3}$\,\% of the S2CLS source flux due to flux boosting.   Whilst these SMGs are faint, we do not see any evidence from their beam-deconvolved continuum sizes that their 870-$\mu$m  emission is more spatially extended than the galaxies detected in the higher resolution maps. This suggests that the missing flux,  at least in these ten maps, is not a result of  flux being resolved out in either our original higher-resolution imaging or the new deeper and lower resolution observations.

\subsection{AGN Fraction}

SMGs have been  proposed as the progenitors for the massive spheroid galaxies seen in the local Universe \citep{lilly1999canada,smail2004rest,simpson2014alma}. Locally such galaxies exhibit a strong correlation between the mass of their central super-massive black holes (SMBH) and the stellar mass of the host galaxy \citep{magorrian1998demography,gebhardt2000relationship,ferrarese2000fundamental,gultekin2009m}. The existence of this correlation has been used to argue that there is some form of co-evolutionary growth of the SMBH and the surrounding host. This suggestion is supported by observations of the star-formation history and AGN activity of the Universe, which both peak at similar redshifts $z\sim $\,2 \citep{connolly1997evolution,merloni2004anti,hopkins2007observational,cucciati2012star,kulkarni2018evolution}.  Support  also comes from simulations of galaxy mergers and AGN activity, which predict that galaxy mergers trigger star formation and then the subsequent fueling of the SMBH creates an AGN which quenches the star formation through feedback winds \citep{hopkins2008cosmological,narayanan2010formation}, consistent  with the proposed evolutionary path of \citet{sanders1988ultraluminous}. Hence surveying the AGN activity in the SMG population not only provides insights into SMBH growth, but also potentially the evolutionary cycle of SMGs.

\subsubsection{X-ray-selected AGN}

The most reliable method to identify AGNs in galaxies is through the detection of luminous X-ray counterparts.   As noted earlier, part of the UDS field has been observed in the X-ray band with {\it Chandra} by the X-UDS survey  \citep{kocevski2018x}. 
X-UDS mapped a total area of 0.33\,deg$^{2}$, of which the  central $\sim$\,100\,arcmin$^{2}$ (coinciding with the 
CANDELS footprint) is three times deeper than the remainder.    A total of 274 SMGs from AS2UDS  are covered by the
X-UDS observations, with 47 of these lying in the deeper CANDELS region.  Considering the high far-infrared luminosities of our SMGs, which is indicative of high star-formation rates (which contributes to the X-ray emission), we adopt a conservative full-band X-ray luminosity limit of $L_{\rm X}\geq$\,10$^{43}$\,erg\,s$^{-1}$ for classification as an AGN, consistent with \cite{franco2018goods}. For consistency, we transform the reported $L_{\rm X}$ from the redshifts quoted in the X-UDS catalogue to those derived from our \textsc{magphys} analysis, where necessary.

Of the 274 SMGs covered by the X-ray observations, just 21 (8\,$\pm$\,2\,\%) are  matched to X-ray counterparts in the X-UDS catalogue based on the positional errors quoted for the X-UDS sources. All of these are classed as AGNs with $L_{\rm X}\geq$\,10$^{43}$\,erg\,s$^{-1}$. 
Of the 47 sources lying within the CANDELS field, only two are matched to X-UDS X-ray sources:  AS2UDS0173.0 and AS2UDS0292.0.

The \textit{Chandra} coverage in the UDS field is relatively shallow for identifying AGN at very high redshifts (c.f.\ 200--600\,ks in UDS, versus 7\,Ms in CDF-S). Hence, we also stacked the X-UDS \textit{Chandra} soft (0.5--2 keV) and hard (2--8 keV) bands at the positions of the  AS2UDS sources which are individually undetected in X-UDS.  To perform this stacking we use  the \textsc{cstack} stacking software\footnote{Developed by T.\ Miyaji. The UDS implementation of \textsc{cstack} will be made public a year after the publication of \citet{kocevski2018x}}. 
We convert the mean stacked count-rates and uncertainties to fluxes using the conversion factors given in \citet{kocevski2018x},
and then to X-ray luminosities by assuming a power-law X-ray spectrum with photon index $\Gamma=$\,1.7. We excluded the 21 SMGs with X-UDS catalogue matches and stacked the remaining 253 SMGs which are individually undetected from the AS2UDS catalogue in three bins of $L_{8-1000\,\mu m}$ (derived from \textsc{magphys}), with the bins chosen to give roughly equal number of sources.   

\begin{figure}
\includegraphics[width=\columnwidth]{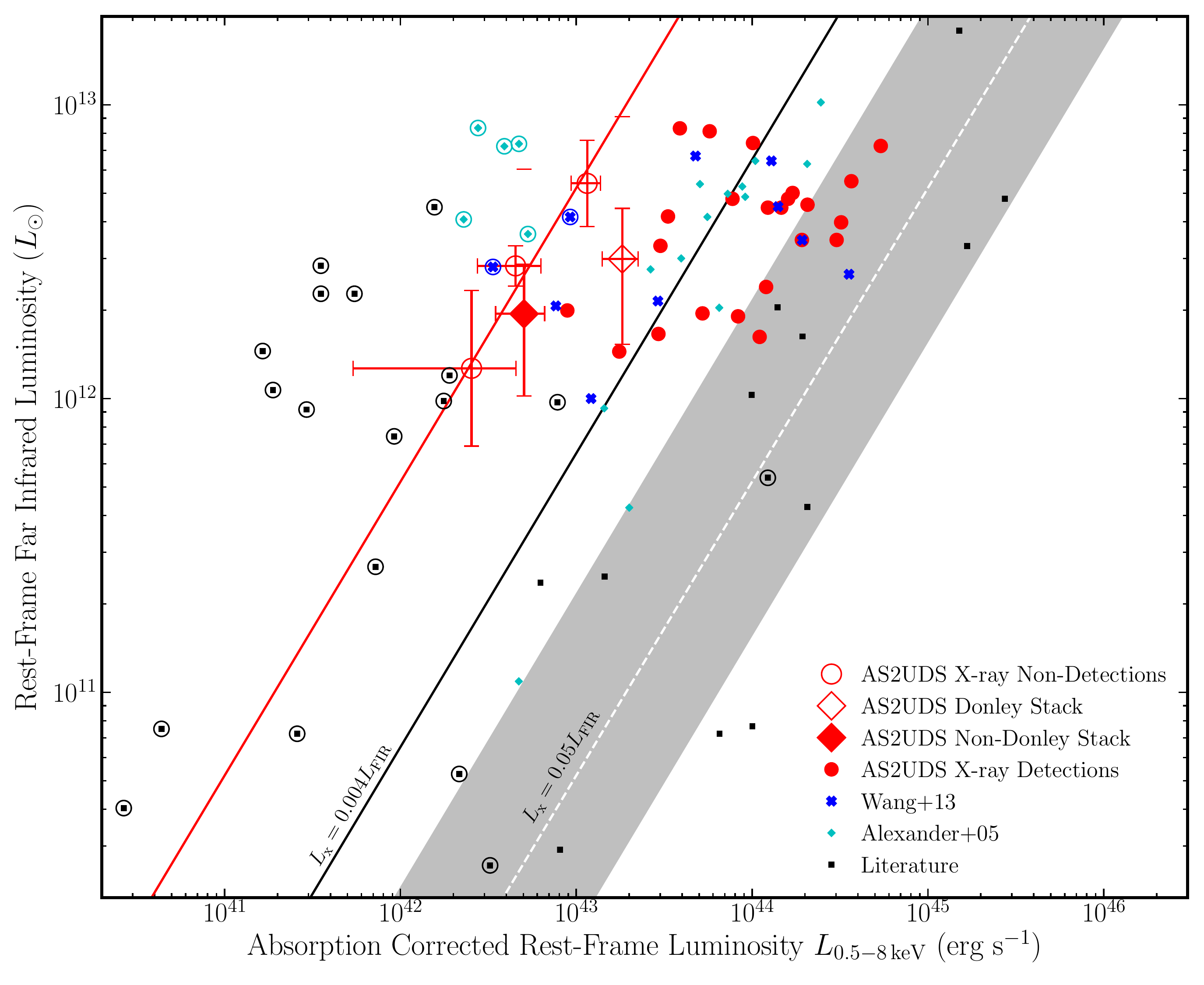}
\caption{We show the relationship between restframe far-infrared (8--1\,000\,\micron) luminosity for the X-ray detected AS2UDS SMGs versus their rest-frame 0.5--8.0\,keV absorption-corrected luminosities. The 21 galaxies with X-UDS matches from our catalogue (red points) and lie  at $L_{\rm x}\gs$\,0.004\,L$_{\rm FIR}$, the approximate dividing line between starburst and AGN dominated galaxies. 
We have derived mean X-ray fluxes by stacking, using  \textsc{CSTACK}, three sub-samples of individually-X-ray-undetected SMGs -- where the sub-samples are ranked on  $L_{\rm FIR}$ -- these are plotted as open symbols with error bars.  These three samples lie  below  the  line dividing starburst and AGN dominated emission.   We conclude that the vast majority of the SMG population, $\gs$\,90\%, do not host luminous AGN.
Galaxies from the literature are plotted as blue crosses \citep{wang2013alma}, cyan diamonds \citep{alexander2005x}, and black squares for galaxies from the literature compiled by \citet{alexander2005x}. Starburst dominated galaxies are denoted by a circle. The white dashed line and grey shaded region are the median luminosity ratio and standard deviation for the quasars from \citet{elvis1994atlas}. The filled diamond is the results of the stacking of the galaxies outside of the \citet{donley2012identifying} selection which show, on average, they are star-formation dominated galaxies whereas the the IRAC selected AGNs in the unfilled diamond appear to be, on average, close to the dividing line between star-formation and AGN dominated.}
\label{fig:xlums}
\end{figure}

We plot the X-ray and far-infrared luminosities of the 21 individually X-ray detected SMGs and the three composite stacked subsamples in Figure~\ref{fig:xlums}.  As noted in \cite{alexander2005rapid} there appears to be a dividing line between AGN-dominated and starburst-dominated galaxies at L$_{\rm X}\sim$\,0.004\,L$_{\rm FIR}$. As expected from the \citet{sanders1988ultraluminous} model of SMG evolution, the SMGs with X-UDS matches cover the entire range of $L_{\rm X}/$\,$L_{\rm FIR}$ from nearly star-formation dominated to the region associated with the AGN-dominated quasars. Nevertheless, the majority of our individually X-ray--detected SMG show $L_{\rm FIR}/L_{\rm X}$ ratios consistent with AGN dominated systems, as expected from their median $L_{\rm X}\sim$\,10$^{44}$\,erg\,s$^{-1}$.   However, the three stacked subsamples of individually X-ray-undetected SMGs all exhibit   ratios of $L_{\rm X}=($\,5$\pm$2\,$)\times$\,10$^{-5} $\,L$_{\rm FIR}$, more consistent with starburst-dominated systems.  This indicates that on-average the X-ray-undetected SMGs are likely to be star-formation dominated and hence the bulk of the SMG population is unlikely to host luminous AGN. 

We estimate the black hole mass accretion rates ($\dot{M}_{\rm acc}^{\rm BH}$) for our SMGs by assuming a $L_{\rm X}$ to $L_{\rm bol}$ bolometric correction factor of 15 \citep{lusso2012bolometric}, an efficiency factor ($\epsilon$) of 0.1 and using Equation \ref{eq:bhacc}

\begin{equation}
   L_{\rm bol}=\epsilon\dot{M}_{\rm acc}^{\rm BH}c^{2}.
	\label{eq:bhacc}
\end{equation}

For our X-ray detected SMGs which we classify as `AGN dominated' this results in $\dot{M}_{\rm acc}^{\rm BH}$ in the range of 0.1--1.4\,$M_{\odot}\,$yr$^{-1}$. For galaxies to evolve along the local relation of spheroid mass-black hole mass, then the ratio of this accretion rate to the star-formation rate must be $\sim$\,2\,$\times$\,10$^{-3}$ \citep{magorrian1998demography,drouart2014rapidly}. Taking the \textsc{magphys} derived star-formation rates, for our AGN dominated SMGs we find a median $\dot{M}_{\rm acc}^{\rm BH}$/SFR ratio of (1.2$\pm$0.5)$\times$10$^{-3}$. This is slightly below the expected ratio,  which may indicate  that these AGN-hosting, but strongly star-forming, SMGs are in a phase where their star-formation rate is exceeding the corresponding SMBH mass growth rate.  However,  the difference is not statistically significant and moreover,  there are significant uncertainties; if the efficiency factor was 0.06 \citep[within the acceptable range of $\epsilon$ e.g.][]{davis2011radiative} then these systems would comfortably lie on the co-evolutionary track of stellar and black hole mass. 

More critically, that is for the relative SMBH and stellar mass growth in the \emph{minority} of SMGs which show AGN signatures.    For the bulk of star-formation-dominated SMGs  (which lie within the X-UDS coverage) we find $\dot{M}_{\rm acc}^{\rm BH}$/SFR ratios of only $\sim$\,0.1\,$\times$\,10$^{-5}$  which is significantly lower than that required for co-evolutionary growth to the local relation.  Adopting physically plausible values for the bolometric correction or efficiency factor cannot change this conclusion. Therefore, for the bulk of  our SMGs to ultimately follow the local relation in spheroid-black hole mass then there must be some subsequent (-or prior) phase of significant growth of the SMBH with comparatively little star-formation, e.g.\ high redshift quasars with their SMBH accretion rates of $\sim$\,1--100\,$M_{\odot}\,$yr$^{-1}$ \citep{hao2008growth}.

\subsubsection{Colour-selected AGN}

In addition to our X-UDS matching and stacking we also supplement our search for AGN activity in the SMG sample by employing IRAC colour-colour selections \citep[e.g.][]{donley2012identifying}.  These use the IRAC bands to identify galaxies with strong power-law emission in the restframe near-infrared, particularly beyond 2\,$\mu$m, which is a good indication of the presence of an AGN, even if it is dust obscured.

\begin{figure*}
\includegraphics[width=2\columnwidth]{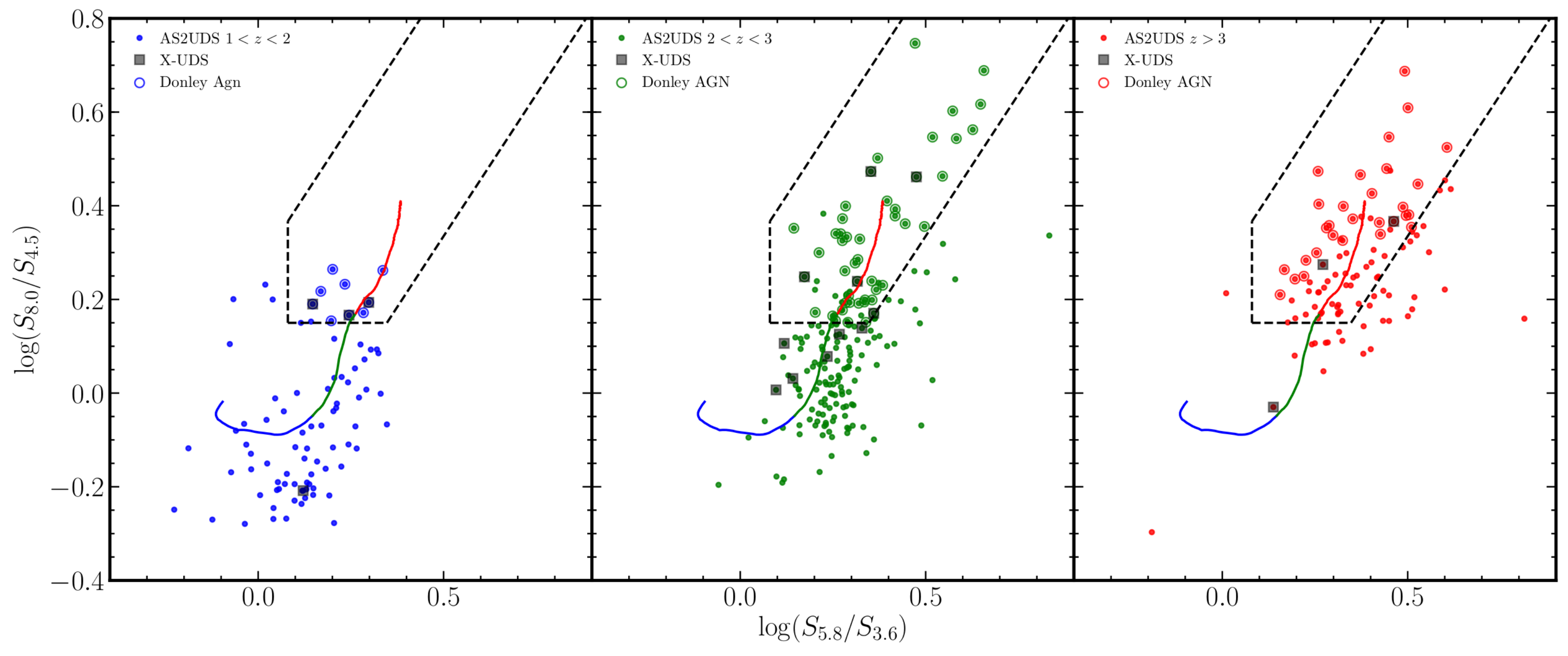}
\caption{We use the IRAC colour--colour selection described in \citet{donley2012identifying} to select candidate AGN from the AS2UDS SMG sample, shown here in three redshift slices. We overlay the predicted colours for the ALESS composite SED \citep{swinbank2014alma} as a function of redshift, colour coded in blue for $z=$\,1--2, green for $z=$\,2--3, and red for $z>$\,3. The track illustrates how above $z\gtrsim$\,3 the AGN colour selection will potentially be  contaminated by star-forming galaxies. The final selection of AGNs based on the `wedge' and the star-formation dominated cut are circled in their respective redshift colours. In total the IRAC-colour selection finds 125 potential AGN candidates across the entire AS2UDS coverage. However, the most reliable  estimate of the AGN fraction comes from the sample lying within the X-UDS region and with redshifts of $z<$\,3, which yields  37 candidates. The 21 X-UDS X-ray-detected SMGs hosting AGN are marked as black squares.}
\label{fig:iracselection}
\end{figure*}

Across the whole of the UDS there are 383/708 SMGs with  coverage in all four IRAC channels (and with photometry that does not suffer from significant contamination from neighbours, see: \S\ref{sec:multi}), necessary to apply the colour selection from \citet{donley2012identifying}.  We show the distribution of these galaxies in    Figure~\ref{fig:iracselection}, but first check that the heavily dust-obscured nature of these sources won't lead to them being misclassified in this colour-colour space.  We therefore plot on Figure~\ref{fig:iracselection} the track of the colours expected from the  composite SED for ALESS SMGs from \citet{swinbank2014alma}.  This shows that at the highest redshifts, $z>$\,3 the typical colours of SMGs will mimic that of an AGN, while at lower redshifts the colours of these obscured and actively star-forming galaxies will fall outside the region populated by AGN. For this reason we apply the additional selection cut from \citet{donley2012identifying} to AS2UDS SMGs in independent redshift slices and we also identify the star-formation dominated high-redshift SMGs which display IRAC colours consistent with a redshifted 1.6\,$\mu$m stellar `bump' and 8.0/3.6\,$\mu$m flux ratios consistent with the local (Ultra)-luminous infrared galaxies SED templates of \cite{rieke2009determining}.

%
%

We test the efficacy of the \citet{donley2012identifying} selection by employing \textsc{cstack} to derive the average X-ray flux of the 25 SMGs with $z_{\rm phot}<$\,3 that are within both the IRAC AGN selection and X-UDS coverage and are not close  to another bright X-ray source, and a control sample  of  131 SMGs outside of the colour selection region. We show these in Figure~\ref{fig:xlums} and see that the 131 colour-selected SMGs lying outside the \citet{donley2012identifying} selection do indeed have a $L_{\rm X}/\,L_{\rm FIR}$ ratio consistent with those expected from star formation. However, the stacked X-ray properties of the 25 SMGs at $z<$\,3 which fall within the \citet{donley2012identifying} selection provide a more ambiguous   $L_{\rm X}/\,L_{\rm FIR}$ ratio, with  these sources lying below the threshold to be classified as an AGN. For this reason we choose to view the IRAC-colour selected samples, even at $z<$\,3, as providing an {\it upper limit} on the potential AGN fraction in SMGs.

\subsubsection{AGN fraction in SMGs}

From the SMGs lying within the X-UDS coverage we derive a lower limit on the X-ray detected AGN fraction of 21/274 (8\,$\pm$\,2\,\%). To estimate an upper limits on the AGN fraction we include the IRAC-colour-selected AGN within this region but employ a $z<$\,3 cut off for reasons discussed above. With the IRAC-selected AGN and the redshift cut we estimate an upper limit of 45/162 (28\,$\pm$\,4\,\%) AGN in the AS2UDS population. 

The range of potential AGN fraction for our flux-limited sample lies in near the middle of results from earlier work in the literature, e.g.\ 38$^{+12}_{-10}$\,\% in \citet{alexander2005x}, (20--29)\,$\pm$\,7\,\% in \citet{laird2010x}, 18$\pm$7\,\% in \citet{georgantopoulos2011x} and $\sim$\,28\,\% in \citet{johnson2013x}, as well as the ALMA-based estimate from ALESS: 17$^{+16}_{-6}$\,\% in \citet{wang2013alma}.

Recently, working with a 1.1-mm selected ALMA sample of SMGs in the GOODS-South field, \cite{franco2018goods} reported a high AGN fraction, $\sim$\,40\,$\pm$\,14\,\%.  Their SMG sample has an 870-$\mu$m equivalent flux range of $S_{870}\sim$\,0.8--3.9\,mJy. To match to this \cite{franco2018goods} selection, we also estimate the AGN fraction for our fainter $S_{870}<$\,4.0\,mJy SMGs and combine the number of our IRAC-selected candidates at $z<$\,3 with the confirmed X-ray bright SMGs from the X-UDS matching to find an upper limit on the AGN fraction of 26\,$\pm$\,5\,\% (28/109) within the X-UDS coverage, consistent with our whole sample. This is lower than the estimate from GOODS-South, but given the significant uncertainty on the latter, we do not give too much weight to this disagreement.

\subsection{Passive Galaxy Progenitors}

SMGs, with their extreme star-formation rates and implied high molecular gas content, could form significant stellar masses ($M_{\ast}$) of 10$^{10}$--10$^{11}$\,M$_{\odot}$ on a timescale of just $\sim$\,100\,Myrs. This rapid formation of a massive system at high redshifts has led to them being proposed as the progenitors of high-redshift compact quiescent galaxies \citep{simpson2014alma,toft2014submillimeter,ikarashi2015compact}, which subsequently evolve into local spheroidal galaxies. Observational support for this evolutionary relation has been claimed from comparisons of the stellar masses of SMGs \citep{swinbank2006link,hainline2011stellar, toft2014submillimeter}, the spatial clustering of relative to that of local ellipticals \citep[e.g.][]{hickox2012laboca,chen2016scuba,wilkinson2016scuba} and the compact rest-frame far-infrared sizes of SMGs \citep{simpson2017scuba,hodge2016kiloparsec}, as well as from theoretical modelling of SMGs and their descendants \citep[][]{gonzalez2011role,mcalpine2019nature}. 

Using our large sample with complete redshift information we can test these claimed connections, especially in light of recent
advances in the studies of high-redshift passive galaxies.  For example \citet{estrada2018clear}, using \textit{HST} grism spectroscopy, have constrained the formation redshift and metallicities for a sample of 32 spectroscopically-classified quiescent galaxies at $z=$\,1--1.8,  which are believed to be massive,  M$_{\ast}>$\,10$^{10}$\,M$_{\odot}$. They find that, nearly independent of the observed redshift of the quiescent galaxies, their \emph{formation} redshift (the epoch where $\gtrsim$\,70\,\% of the stellar mass has already formed) is $z_{\rm form}>$\,2--3, with a third of their sample having $z_{\rm form}>$\,3. Constraints on their metallicities suggesting that these star-forming progenitors must have already enriched to approximately Solar metallicities (which is consistent with the high dust masses of the SMGs as well as the crude estimates of their metallicities from \citet{swinbank2004rest,takata2006rest}). \citet{morishita2018massive}, likewise, looked at the mass accumulation and metallicity history for a sample of 24 apparently massive galaxies ($M_{\ast}>$\,10$^{11}$\,M$_{\odot}$) at $z=$\,1.6--2.5 via SED modelling and inferred that the majority of their sample had formed $>$\,50\,\% of their mass around $\sim$\,1.5\,Gyr prior to their observed redshifts, yielding formation ages similar to \citet{estrada2018clear}.

\begin{figure}
\includegraphics[width=\columnwidth]{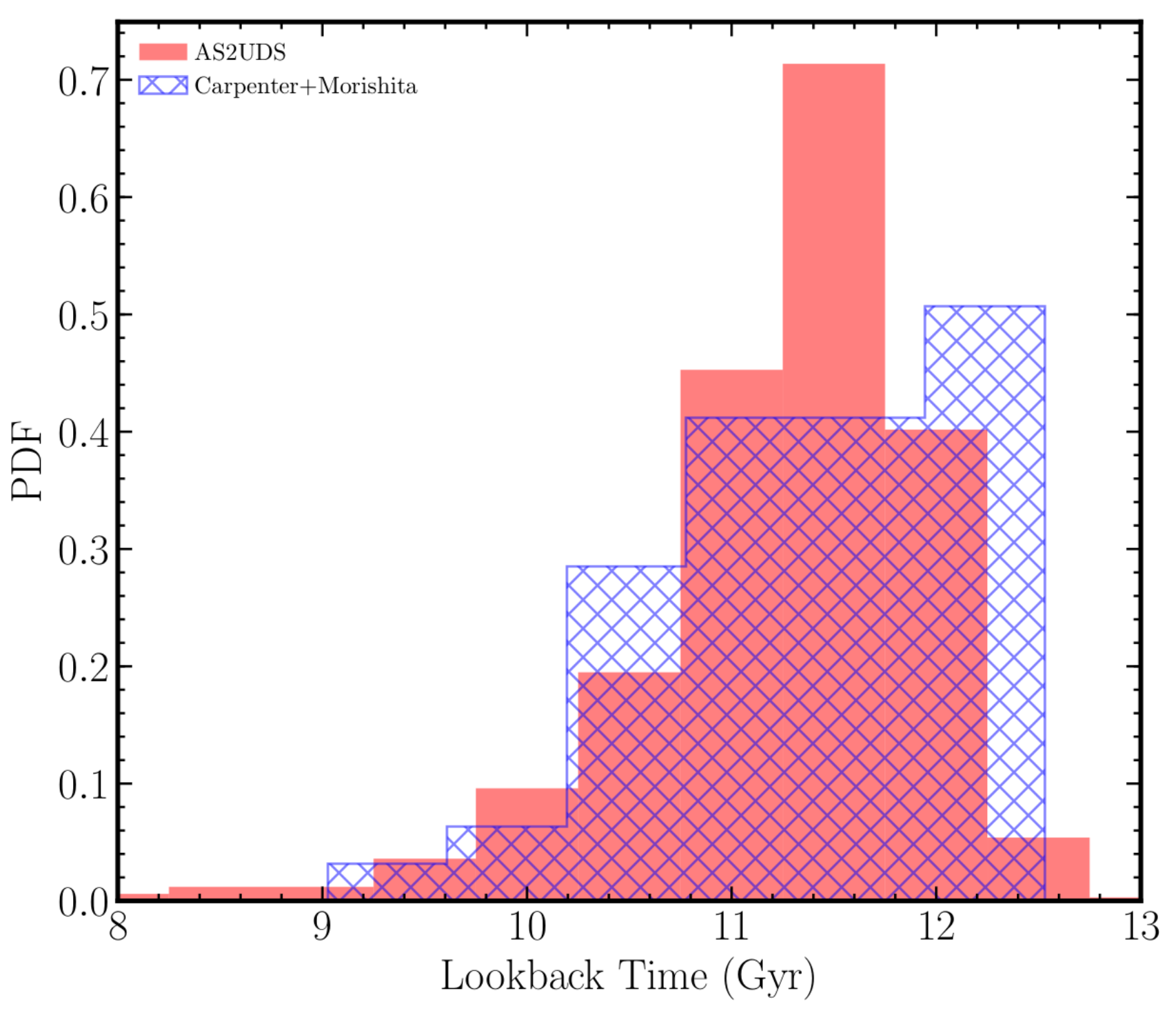}
\caption{The inferred mass-weighted ages of the AS2UDS SMGs compared to the \emph{formation} redshift for  $z\sim$\,1--2.5 passive galaxies from \citet{estrada2018clear} and \citet{morishita2018massive}. The high-redshift quiescent galaxy population across this broad redshift range are found to have similar \emph{formation} redshifts, which in turn broadly match the redshift distributions for the formation of the SMG galaxy population.  This is consistent with the interpretation of the SMGs as  likely progenitors for the spectroscopically confirmed quiescent galaxies at $z\sim$\,1--2.5.}
\label{fig:carpenter}
\end{figure}

In Figure~\ref{fig:carpenter} we plot the age distribution for massive AS2UDS SMGs and the combined \citet{estrada2018clear} and \citet{morishita2018massive} \emph{formation} redshifts for their samples of  quiescent galaxies.  We adopt a mass limit of M$_{\ast}>$\,10$^{9.85}$\,M$_{\odot}$, from our {\sc magphys} estimated stellar masses to match the passive galaxy samples, this accounts for some continuing star-formation activity and associated stellar mass growth in these systems, although our conclusions are not sensitive to this assumption. For a more accurate comparison of `formation' ages we derive the mean offset from the observed redshift of our SMGs to the `age' of their mass-weighted stellar population. This offset is calculated from the inferred gas masses, star-formation rates, and stellar masses from {\sc magphys} for the AS2UDS sample. We estimate (and apply) a typical +200\,Myr offset to our observed redshifts to correct to a mass-weighted stellar population age which is more representative of the period in which the galaxy was forming the majority of its stellar mass.

Figure~\ref{fig:carpenter} shows that the AS2UDS age distribution is comparable to the formation ages which are inferred for massive, quiescent galaxies at $z\sim$\,1--2.5: both distributions peak at lookback times of 11.5--12.5\,Gyr, median ages for the distributions are in agreement with 11.4$^{+0.1}_{-0.2}$\,Gyr for AS2UDS and 11.5\,$\pm$\,0.3\,Gyr for the comparison sample, and both populations show a younger stellar population $<$\,11\,Gyr tail containing $\sim$\,30\,\% of their respective distributions.   This provides further support for the claims that
the SMG and high-redshift massive quiescent galaxy populations may have an evolutionary link.

We can also ask if the number density of the SMGs and high-redshift passive galaxies are consistent with any likely evolutionary cycle. We can derive a co-moving number densities for the subset of the quiescent population from \citet{estrada2018clear} with formation redshifts $z_{\rm form}=$\,2--3 (the population where number densities are available), which is $n_{\rm qg}\sim $\,2.5$\times10^{-4}$\,Mpc$^{-3}$. This compares to a number density of $n_{\rm SMG}=($\,2.8$^{+0.2}_{-0.1})\times$\,10$^{-5}$\,Mpc$^{-3}$ for the SMGs in the same redshift range after applying the $M_{\ast}>$\,10$^{9.85}$\,M$_{\odot}$ mass cut. The SMG number density is 11$^{+3}_{-2}$\,\% of the quiescent galaxy number density estimated by \cite{estrada2018clear}. However, we also need to correct the apparent SMG number density for a duty cycle as they are thought to attain a
high star formation rate for a comparatively short duration, compared to the redshift range being considered.

The lifetime of the high-SFR phase of SMGs is dependant on either a simple gas depletion timescale, or through some star formation quenching mechanism e.g.\ active galactic nuclei feedback. Previous estimates of the lifetimes of the SMG-phase based on gas depletion timescales or clustering analysis have suggested durations of $\sim$\,40--200\,Myr \citep{tacconi2006high, swinbank2006link, riechers2011dynamical, hickox2012laboca, bothwell2013survey}. We can estimate this average duty cycle duration using a simple model of the SMG evolution by making the following assumptions: (i) the $z=$\,2--3, $M_{\ast}>$\,10$^{9.85}M_{\odot}$ AS2UDS SMGs are progenitors of the $z=1$--1.8 quiescent galaxies and likewise all $z=$\,1--1.8 quiescent galaxies are the descendants of these SMGs, (ii) each SMG has a single burst of intense star-formation. With these assumptions the burst duration can be estimated by 
\begin{equation}
    t_{\rm burst}=t_{\rm obs}\times\bigg(\frac{n_{\rm SMG}}{n_{\rm qg}}\bigg),
\end{equation}
where $t_{\rm burst}$ is the burst duration, $t_{\rm obs}$ is the duration of the epoch we calculated the SMG comoving density ($z=$\,2--3), and $n_{\rm SMG}/n_{\rm qg}$ are the co-moving number densities for the SMGs and quiescent galaxies respectively. For our measured number density we find a burst duration of 190$^{+50}_{-40}$\,Myr which is consistent with the other estimates derived from observed gas masses and star-formation rates of SMGs within this redshift range in the literature \citep{hickox2012laboca, bothwell2013survey}. Therefore the space-density of dusty star-forming galaxies at $z=$\,2--3 that form the bulk of the AS2UDS sample, are consistent with that required for them to comprise  the progenitors of quiescent galaxy population seen at $z=$\,1--1.8.
\section{Conclusions} \label{sec:conclusions}

We have presented the catalogue for the largest  homogeneously-selected sample of sub-millimetre galaxies to date, an ALMA 870\,$\mu$m continuum follow-up survey of  716, $>4$\,$\sigma$ single-dish sub-millimetre sources selected from the SCUBA-2 Cosmology Legacy Survey 850-$\mu$m map of the UKIDSS UDS field. Our deep, high-resolution ALMA observations identified 708 $>$\,4.3\,$\sigma$ sources which account for the majority of the flux detected in the parent SCUBA-2 map.  The main conclusions of this study are:

\begin{itemize}

\item Utilising the extensive multi-wavelength coverage of the UDS field we fit SEDs for our galaxies using \textsc{magphys} and from these fits derived a median photometric redshift for our galaxies of $z_{\rm phot}=$\,2.61\,$\pm$\,0.09 with a high-redshift tail comprising 33$^{+3}_{-2}$\,\% of SMGs with $z_{\rm phot}>$\,3. 

\item From the subset of SMGs with CANDELS imaging we find that 50$\pm$10\,\% show either clear merger morphologies or have likely companions, displaying similar colours, on $<20$\,kpc scales. These likely interacting systems have a median redshift at $z_{\rm phot}=2.2\pm0.1$, which is significantly lower than the median photometric redshift. When we select SMGs with redshifts $z_{\rm phot}<2.75$, to account for the redshift at which we reasonably expect to reliable detect interactions in the CANDELS imaging, then this `likely interacting' fraction accounts for $\sim$80\,\% of SMGs. This suggests that the elevated star-formation rates in these systems are driven by mergers.

\item With our large sample size we see convincing evidence for  evolution in the $S_{870}$ flux density of  sources with redshift with a best fit trend gradient of 0.09\,$\pm$\,0.02\,mJy$^{-1}$. This evolution was not robustly identified in  previous smaller  surveys due to their limited statistics and we show how reducing our sample size down to $\sim$\,100 galaxies results in a statistically insignificant result. The consequence of this trend is that on average our most luminous galaxies are found at higher redshifts in comparison to less active galaxies, a strong indication of galaxy downsizing.

\item Through stacking \textit{Herschel} observations at the positions of the 101 SCUBA-2 sources for our ALMA maps produced no $>$\,4.3-$\sigma$ detections, we show that these sources are not dominated by false-positive detections in the parent S2CLS catalogue. We find an overdensity of, on average, $\sim$\,1.5 $K$-band sources at the locations of these `blank' ALMA maps at redshifts $z=$\,1.5--4 which, combined with the strong evidence from SPIRE-stacking that the original SCUBA-2 flux is real,  suggests that the lack of ALMA counterparts is a result of blending of the sub-millimetre emission from $\gs$\,1--2  \emph{faint}  galaxies at these positions. We confirm that this is the case with deeper repeat Cycle 5 observations of ten examples of these `blank' maps which yield 16 new ALMA detections below our previous flux limit. This has significant consequences for the faint-end number counts of sub-millimetre galaxies.

\item We identify AGNs associated with our SMG sample by both matching our catalogue to the X-UDS {\it Chandra} X-ray coverage of the field and also by applying an IRAC colour-colour selection. We estimate a lower limit on our AGN fraction from the X-ray detections of 8\,$\pm$\,2\,\% and an upper limit by including our IRAC-colour selected AGNs of 28\,$\pm$\,4\,\%. This range is consistent with previous results, although somewhat lower than the most recent results reported for a small sample in the GOODS-S field.   We conclude that most sub-millimetre bright galaxies do not host an unobscured or moderately obscured luminous AGN.

\item Looking to the most likely candidate descendants for our SMGs we compare the constraints on their  redshift and number density (as well as stellar mass and metallicity) to those   expected for the progenitors of $z=$\,1--2.5 quiescent galaxies predicted by \citet{estrada2018clear} and \citet{morishita2018massive}. We find that the properties of the AS2UDS SMG population  are consistent with these constraints, with median mass-weighted ages for the SMGs of 11.4$^{+0.1}_{-0.2}$\,Gyr, in good agreement with the  median formation ages for the quiescent galaxies of 11.5$\pm0.3$\,Gyr.The number density  of the SMGs and \citet{estrada2018clear} populations are also consistent with this evolutionary link if 
the typical star-formation burst  duration of the SMGs is $\sim$\,190$^{+50}_{-40}$\,Myr, which is similar to previous independent estimates.

\end{itemize}

\section*{Acknowledgements}
SMS\ acknowledges the support of STFC studentship (ST/N50404X/1). AMS\ and IRS\ acknowledge financial support from an STFC grant (ST/P000541/1). IRS\, EAC\, and BG\ also acknowledge support from the ERC Advanced Investigator program DUSTYGAL (321334). JLW acknowledges support from an STFC Ernest Rutherford Fellowship (ST/P004784/1 and ST/P004784/2). JEG acknowledges support from a Royal Society University Research Fellowship. MJM acknowledges the support of the National Science Centre, Poland through the POLONEZ grant 2015/19/P/ST9/04010; this project has received funding from the European Union's Horizon 2020 research and innovation programme under the Marie Sk{\l}odowska-Curie grant agreement No. 665778. T. Miyaji and the development of CSTACK is supported by UNAM-DGAPA IN104216,IN111319 and CONACyT 252531. This work makes use of CSTACK (http://cstack.ucsd.edu/ or http://lambic.astrosen.unam.mx/cstack/) developed by Takamitsu Miyaji. The ALMA data used in this paper were obtained under programs ADS/JAO.ALMA\#2012.1.00090.S, \#2015.1.01528.S and \#2016.1.00434.S. ALMA is a partnership of ESO (representing its member states), NSF (USA) and NINS (Japan), together with NRC (Canada) and NSC and ASIAA (Taiwan), in cooperation with the Republic of Chile. The Joint ALMA Observatory is operated by ESO, AUI/NRAO, and NAOJ.  This paper used data from project MJLSC02 on the James Clerk Maxwell Telescope, which is operated by the East Asian Observatory on behalf of The National Astronomical Observatory of Japan, Academia Sinica Institute of Astronomy and Astrophysics, the Korea Astronomy and Space Science Institute, the National Astronomical Observatories of China and the Chinese Academy of Sciences (Grant No.\ XDB09000000), with additional funding support from the Science and Technology Facilities Council of the United Kingdom and participating universities in the United Kingdom and Canada. UKIDSS-DR11 photometry made use of UKIRT. UKIRT is owned by the University of Hawaii (UH) and operated by the UH Institute for Astronomy; operations are enabled through the cooperation of the East Asian Observatory. When (some of) the data reported here were acquired, UKIRT was supported by NASA and operated under an agreement among the University of Hawaii, the University of Arizona, and Lockheed Martin Advanced Technology Center; operations were enabled through the cooperation of the East Asian Observatory. When (some of) the data reported here were acquired, UKIRT was operated by the Joint Astronomy Centre on behalf of the Science and Technology Facilities Council of the U.K.


\bibliographystyle{mnras}
\bibliography{AS2UDS_Cat2}




\bsp	
\label{lastpage}
\end{document}